\documentclass[twocolumn,apj]{emulateapj}
\usepackage{graphicx}
\usepackage{rotating}
\usepackage{longtable, lscape}

\slugcomment{To appear in The Astrophysical Journal}

\shorttitle{Single-lobed Stokes V profiles in quiet sun}
\shortauthors{Sainz Dalda, A., Mart\'inez-Sykora, J., Bellot Rubio, L., Title, A.} 

\newcommand{\tresfig}[7]
{\begin{figure} 
\centering
 \begin{minipage}[c]{0.3\textwidth}
  \centering \includegraphics[#1]{#2}
 \end{minipage}%
 \begin{minipage}[c]{0.3\textwidth}
  \centering \includegraphics[#3]{#4}
 \end{minipage}
 \begin{minipage}[c]{0.3\textwidth}
  \centering \includegraphics[#5]{#6}
 \end{minipage}
\caption{#7}
\end{figure}}

\newcommand{\unafig}[3]
{\begin{figure} 
\includegraphics[width=0.5\textwidth]{#2}
\caption{#3}
\end{figure}}

\newcommand{\dosfig}[7]
{\begin{figure*} 
  \includegraphics[width=0.5\textwidth]{#2}
  \includegraphics[width=0.5\textwidth]{#4}
\caption{#5}
\end{figure*}}

\newcommand{\bop}{blue-only profile}
\newcommand{\rop}{red-only profile}
\newcommand{\sls}{single-lobed Stokes V profile}

\begin{document}
\title{Study of single-lobed circular polarization profiles in the quiet sun}
\author{A. Sainz Dalda\altaffilmark{1}, J. Mart\'inez-Sykora\altaffilmark{2,3}, L. Bellot Rubio\altaffilmark{4}, A. Title\altaffilmark{3}}
\affil{$^1$Stanford-Lockheed Institute for Space Research, Cypress Hall, 466 Via Ortega, Stanford, CA 94305-4085, USA}
\affil{$^2$Lockheed Martin Solar and Astrophysics Laboratory, 3176 Porter Dr., Palo Alto, CA 94304}
\affil{$^3$Institute of Theoretical Astrophysics, University of Oslo, P.O. Box 1029 Blindern, N-0315 Oslo, Norway}
\affil{$^4$Instituto de Astrof\'isica de Andaluc\'ia, CSIC, Apdo. 3004, 18080 Granada, Spain}
\email{asdalda@stanford.edu or asainz@lmsal.com}

\clearpage

\begin{abstract}
The existence of asymmetries in the circular polarization (Stokes V) 
profiles emerging from the solar photosphere is known since the 1970s. 
These profiles require the presence of a velocity gradient along the line of 
sight, possibly associated with gradients of magnetic field strength, 
inclination and/or azimuth. 
We have focused our study on the
Stokes V profiles showing extreme asymmetry in the from of only one lobe. 
Using Hinode spectropolarimetric measurements we have performed a 
statistical study of the properties of these profiles in the quiet sun. We 
show their spatial distribution, their main physical properties, 
 how they are related with several physical observables and their 
behavior with respect to their position on the solar disk. The single lobed 
Stokes V profiles occupy roughly 2\% of the solar surface. For the first time, 
we have observed their temporal evolution and have retrieved the physical 
conditions of the atmospheres from which they emerged using an inversion 
code implementing discontinuities of the atmospheric parameters along the 
line of sight.  In addition, we use synthetic Stokes profiles from 3D 
magnetoconvection simulations to complement the results of the inversion. The main 
features of the synthetic single-lobed profiles are in general agreement 
with the observed ones, lending support to the magnetic and dynamic topologies 
inferred from the inversion. 
The combination of all these different analysis suggests that most of the single-lobed
Stokes V profiles are signals coming from magnetic flux emergence and/or submergence
processes taking place in small patches in the photospheric of the quiet sun.

\end{abstract}


\clearpage

\section{Introduction}
Stokes parameters are used to characterize the polarization state of light: 
the Stokes I parameter tells us about the intensity, Q and U about the linear 
polarization and V about the circular polarization \citep{Ree89, Lan92, del03book}. 
The variation of these parameters across a spectral line are called Stokes profiles. 
When the light beam comes from a static atmosphere, the Stokes I, Q and U 
profiles are symmetric functions, and Stokes V is antisymmetric showing two
 lobes of opposite signs with respect to the central wavelength.
However, the Stokes V signals coming from the solar atmosphere rarely are  
purely antisymmetric, i.e., the two lobes are different either in shape, in 
amplitude or in area. The asymmetry of the  Stokes V is generally 
described in terms of the difference between the amplitudes (or the areas) 
of each lobe normalized to their sum\footnote{Therefore, the area 
asymmetry is in some way the {\it net circular polarization} normalized 
to the total area.}. 
The asymmetry in the Stokes V profile requires a velocity 
gradients in the atmosphere where the radiation is coming from 
(\citealt{Ill75, Aue78, Lan96} among others). In case of extreme asymmetries, 
one of the lobes of Stokes V is missing. These are referred as \sls s. 
Such large asymmetries cannot be explained with a velocity gradient alone, 
and require gradients of the magnetic field vector.

In this paper we present a detailed analysis of the single-lobed Stokes V 
profiles observed in the quiet sun. These profiles show only one lobe, or 
two lobes with one reduced by a very large extent. Profiles of this 
type were first described by \cite{Sig99a} and studied in detail by \cite{Sig01} and \cite{Gro00}. 
The former author studied the amount of blue-only and red-only single-lobed 
Stokes V profiles in the quiet sun\footnote{For convenience, we refer to the 
blue-only single-lobed Stokes V profiles as {\it blue-only profiles} and 
the locations where they form clumps as {\it blue-only patches}. Similarly, 
for the red-only single-lobed Stokes V profiles we use the terms {\it red-only 
profiles} and {\it red-only patches}.}. \cite{Sig01} observed a ratio of blue-only to 
red only profiles of 4:1. He associated the blue-only profiles with magnetic 
structures exhibiting downflows whereas the red-only profiles seemed to be 
related to upflows. He used a large scan area, so he could not observed the 
temporal evolution of these signals. Therefore, although he 
proposed several scenarios, he could not explain the phenomena behind them. 

To explain the existence of Stokes V profiles with strong asymmetry several theoretical models 
featuring large gradients in velocity and magnetic field have been suggested.
If the stratification is simple enough, the blue-only lobe in the Stokes V profile 
is stronger (weaker) than the red lobe if the LOS velocity gradient as a function of 
depth is opposite (equal) to the magnetic field strength gradient \citep{Ill75,Aue78}.
\cite{Ste00} proposed that single-lobed profiles can be produced when the line 
of sight passes through a magnetopause, i.e., 
a separatrix layer dividing the atmosphere vertically in two regions with different 
magnetic fields. Magnetopauses occur frequently in the solar atmosphere. 
Examples are the canopies created by expanding flux tubes in the network 
and the interfaces between different magnetic components in sunspot 
penumbrae \citep[see][]{Gro00, Ste00, Schl98}. \cite{Ste00} 
suggested  that a tiny loop expanding towards the upper 
layers, being smaller than the spatial resolution, could create \sls s if 
the expansion of the magnetic field lines in one of the footpoints of the 
loop is larger than in the other. In this case, the footpoint related with 
the large expansion would create asymmetric Stokes V profiles (without 
extreme asymmetry), while the footpoint with normal 
expansion would produce symmetric Stokes V profiles of opposite sign. 
The superposition of both profiles - as a result of limited spatial resolution - 
would cancel one of the lobe and, therefore, the observed profile would be a 
\sls (see Figure~9 in \citealt{Ste00}). \cite{San96} synthesized 
\sls s under the MISMA assumption. 
This type of unusual Stokes V profiles has been related with upward traveling 
shocks driven by an aborted convective collapsed \citep{3lobe_Bel01}. 
They have been observed in convective collapse events that take place 
simultaneously in the photosphere and chromosphere \citep{3lobe_Fis09}. 
\cite{3lobe_Nar10} found that Stokes V profiles with large amplitude asymmetries
are exclusively located at the boundary between regions with 
small-scale abnormal granulation and much larger normal-looking granules. 
Recently, \cite{3lobe_Vit11} have successfully inverted \sls s located either 
in the neutral line of opposite polarity patches or in the {\it very} quiet sun.

The identification of the physical mechanism(s) behind the single-lobed 
circular polarization profiles observed in the quiet sun represents both an 
observational and a theoretical challenge. Here, we use all the tools at 
our disposal to shed light on these intriguing profiles. We analyze a large 
number of spectropolarimetric measurements taken by Hinode at high 
spatial resolution in order to determine the observational properties of the 
profiles in a statistically significant way (Sections~\ref{sec_datos}-\ref{sec_solardisk})
and to investigate their temporal evolution (Section~\ref{sec_temp}). 
We also apply Stokes 
inversions to the data based on models with discontinuities along 
the line of sight (Section~\ref{sec_inv}). The inversions successfully reproduce 
the observed spectra. Finally, we search for single-lobed Stokes V 
profiles in 3D magnetoconvection simulations of the quiet sun and 
study the magnetic and dynamic topologies associated with them. 
Our combined approach allows us to draw a general picture of this 
phenomenon. A complementary analysis of single-lobed profiles in 3D 
magnetoconvection simulations has been performed recently by 
\citet{3lobe_Vit12}.

\section{Data} \label{sec_datos}
The data used in this paper were obtained by the Spectro-Polarimeter 
(SP) located in the Focal Plane Package (FPP) of the Solar Optical 
Telescope (SOT, \citealt{iTsu08}) on board the Hinode satellite \citep{iKos07}. 
This instrument, hereafter referred to as Hinode SOT/SP measures the 
four Stokes parameters of the \ion{Fe}{1} lines at 6301 and 
6302 \AA\  with spectral sampling of 21.5 $m$\AA\ .
The spatial scale can be 0.15\arcsec\ or 0.30\arcsec\ for the X direction 
(perpendicular to the slit direction) and 0.16\arcsec\ or 0.32\arcsec\ for the Y 
direction (parallel to the slit). For the first time,  these data allow to investigate 
the four Stokes parameters of the \sls s under seeing-free conditions and with 
unmatched spatial resolution.

We use 72 data sets spanning a wide range of observational conditions:
exposure time (1.6, 3.2, 4.8, 8.0, 9.6 and 12.6 s); signal to noise ratio 
(between 500 and 1000, calculated as the inverse of the standard 
deviation\footnote{The standard deviation, $\sigma$,  is calculated over 
11 Stokes V maps normalized to the intensity of the continuum
from the first 11 spectral positions in the 
continuum of the observed profile, i.e. from 6300.90 \AA\ to 6301.13 \AA.}, 
$\sigma$);  position on the solar disk (56  maps with 1.00 $> \mu >$ 
0.75, 8 maps with 0.75 $> \mu >$ 0.50 and 8 maps with 
0.50 $> \mu >$ 0); and spatial scale (as mentioned above). 
Our data sets contain scans of large quiet sun regions and 
narrower scans with typical widths of 2-10 granules, for a variety of 
heliocentric distances. Table \ref{latabla} in the appendix summarizes
the basic parameters of the different observing runs. All the analyzed 
profiles have been calibrated using {\it Solarsoft} routines.

\section{Data Analysis}\label{sec_analisis}
To identify single-lobed profiles, we have developed a code that 
checks the shape of the Stokes V spectra and flags the ones that 
have only one lobe. This code was applied to almost $5.2\times10^{7}$ 
pixels corresponding to the 72 selected observations. The code looks 
into the shape of the Stokes V profiles, and  takes into account the following constraints:  
\begin{itemize}
\item[-] The maximum (minimum) signal must be greater (smaller) than 4 (-4) 
times $\sigma$.
\item[-] At least three wavelength points must show signals above (below) the 
4$\sigma$ (-4$\sigma$) threshold.
\item[-] Simultaneously, the other lobe and the minimum (maximum) of 
the whole Stokes V profile must be greater (less) than -3$\sigma$ (3$\sigma$).
\item[-] In the continuum, the circular polarization signal can never 
exceed $\pm$ 3$\sigma$.
\end{itemize}
If these conditions are satisfied by one of the lobes of \ion{Fe}{1} 6301 \AA\ or 6302 \AA\, 
then the profile is flagged as a single-lobed profile. 
For \ion{Fe}{1} 6301 \AA, the blue lobe goes from 
6301.18$\pm0.02$ \AA\ ($\lambda_{b}$)\footnote{For clarity 
in the text we removed the uncertainty ($\pm 0.02$) in the 
measurement of the spectral line positions 
but it is present and the same in all the observational data 
studied in this work unless noted.} to 
6301.50 \AA\ ($\lambda_{c}$),
and the red lobe from 6301.50 \AA\ to 
6301.82 \AA\ ($\lambda_{r}$). For \ion{Fe}{1} 6302 \AA, 
the blue lobe is between 6302.17 \AA\ ($\lambda_{b}$) 
and 6302.49 \AA\ ($\lambda_{c}$), while the red one 
goes from 6302.49 \AA\ to 6302.82 \AA ($\lambda_{r}$).

In order to relate the location of the \sls s with physical parameters, 
we have extracted several line parameters 
from the observed Stokes spectra: 
\begin{itemize} 
\item[-] Stokes I map: for the slit reconstructed Stokes I map we 
used 8 spectral positions on the continuum, located between 
6303.12 \AA\ and  6303.27 \AA.
\item[-] Stokes V map: in this case we used 36 spectral positions located 
on the blue wing of the Stokes V profiles of \ion{Fe}{1} 6302 \AA, roughly from 
6301.72 \AA\ to  6302.49 \AA.
\item[-] Doppler velocities, determined through line bisectors. We use the 
line bisector at the 70\% intensity level (the midpoint between the flanks 
of the line at that intensity) minus the line core wavelength position of the 
average quiet sun profile. The difference is converted into velocity and 
expressed in km s$^{-1}$. Thus, the zero velocity is given by the local average 
 quiet sun profile  
\item[-]  The mean circular polarization degree is calculated as:
$MCPD = (\int_{\lambda_{0}}^{\lambda_{1}}{\frac{|V(\lambda)|d\lambda}{I(\lambda)}})
        /(\lambda_{1} - \lambda_{0})$
\item[-]  The mean linear polarization degree is calculated as:
$MLPD = (\int_{\lambda_{0}}^{\lambda_{1}}{\frac{\sqrt{Q^2(\lambda) + 
        U^2(\lambda)}}{I(\lambda)}d\lambda})/(\lambda_{1} - \lambda_{0})$
\item[-]  The mean total polarization degree is calculated as:
$MTPD = (\int_{\lambda_{0}}^{\lambda_{1}}{\frac{\sqrt{Q^2(\lambda) +
        U^2(\lambda) + V^2(\lambda)}}{I(\lambda)}d\lambda})
        /(\lambda_{1} - \lambda_{0})$
\end{itemize}
The integrals are evaluated from  $\lambda_{0}$=6301.29 \AA\ to 
$\lambda_{1}$=6301.71 \AA\ in the case of 
 \ion{Fe}{1} 6301 \AA, and from $\lambda_{0}$=6302.27 \AA\  
 to $\lambda_{1}$=6302.70 \AA\ for \ion{Fe}{1} 6302 \AA. 
These wavelength intervals cover all the relevant polarization signals, 
avoiding the contribution of the continuum. 

\section{Spatial distribution of single-lobed Stokes V profiles} \label{sec_spatial}
In Figure 1 we show examples of blue-only and red-only Stokes V profiles 
(left and right panels, respectively). There are clear Stokes Q and U 
signals associated with the blue-only profiles, indicating the existence 
of inclined magnetic fields with respect to the line of sight. Furthermore, 
both Q and U are asymmetric profiles. For the red-lobe profile the Stokes 
Q and U signals are much smaller and appear to consist of a single lobe on 
the red wing of the lines. The continuum intensities are larger than 1 and 
around 1 for the blue-only and red-only profiles, respectively. This means 
that the blue-only profile corresponds to a bright structure in the intensity map.
 
\dosfig{width=7cm}{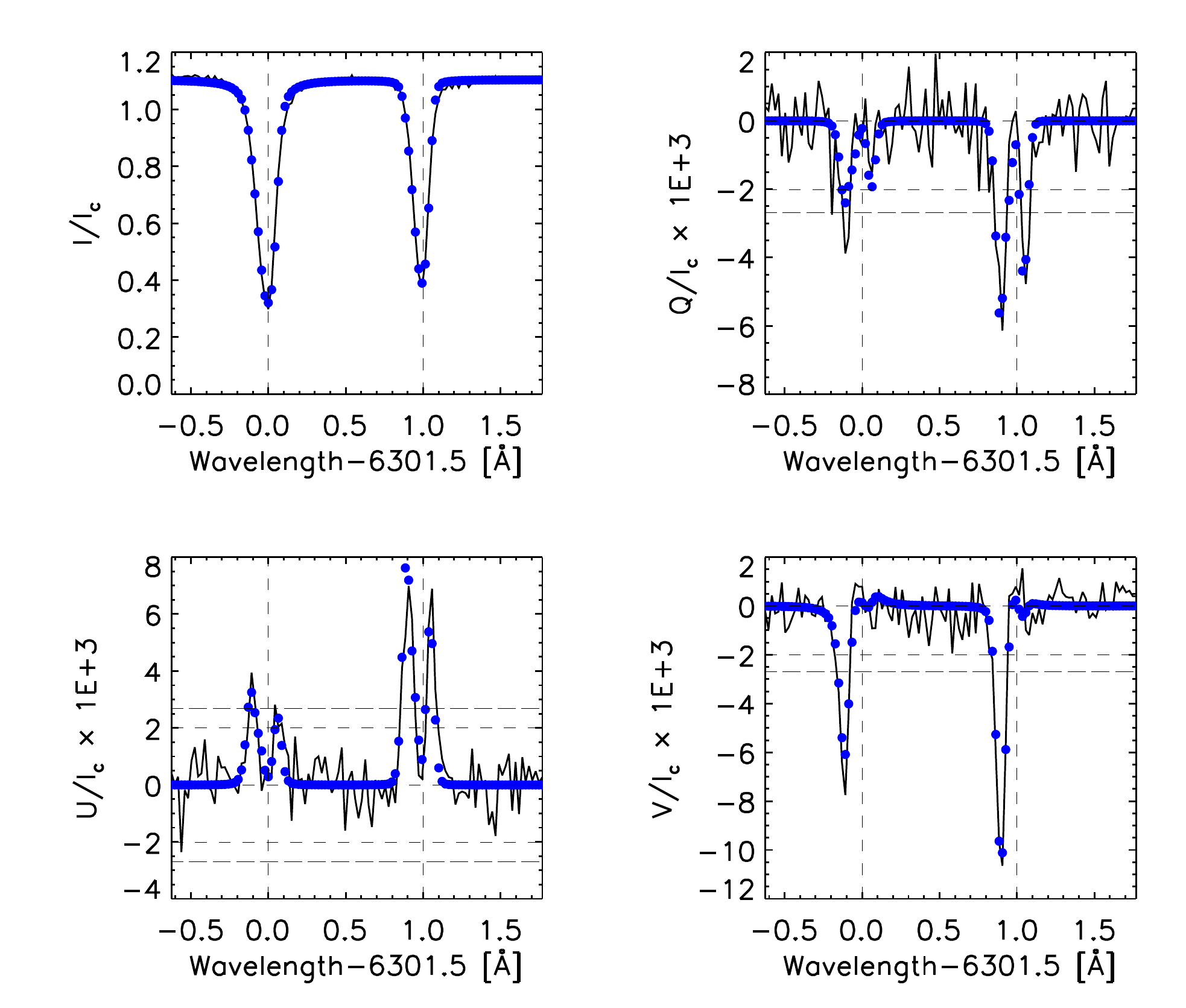}
       {width=7cm}{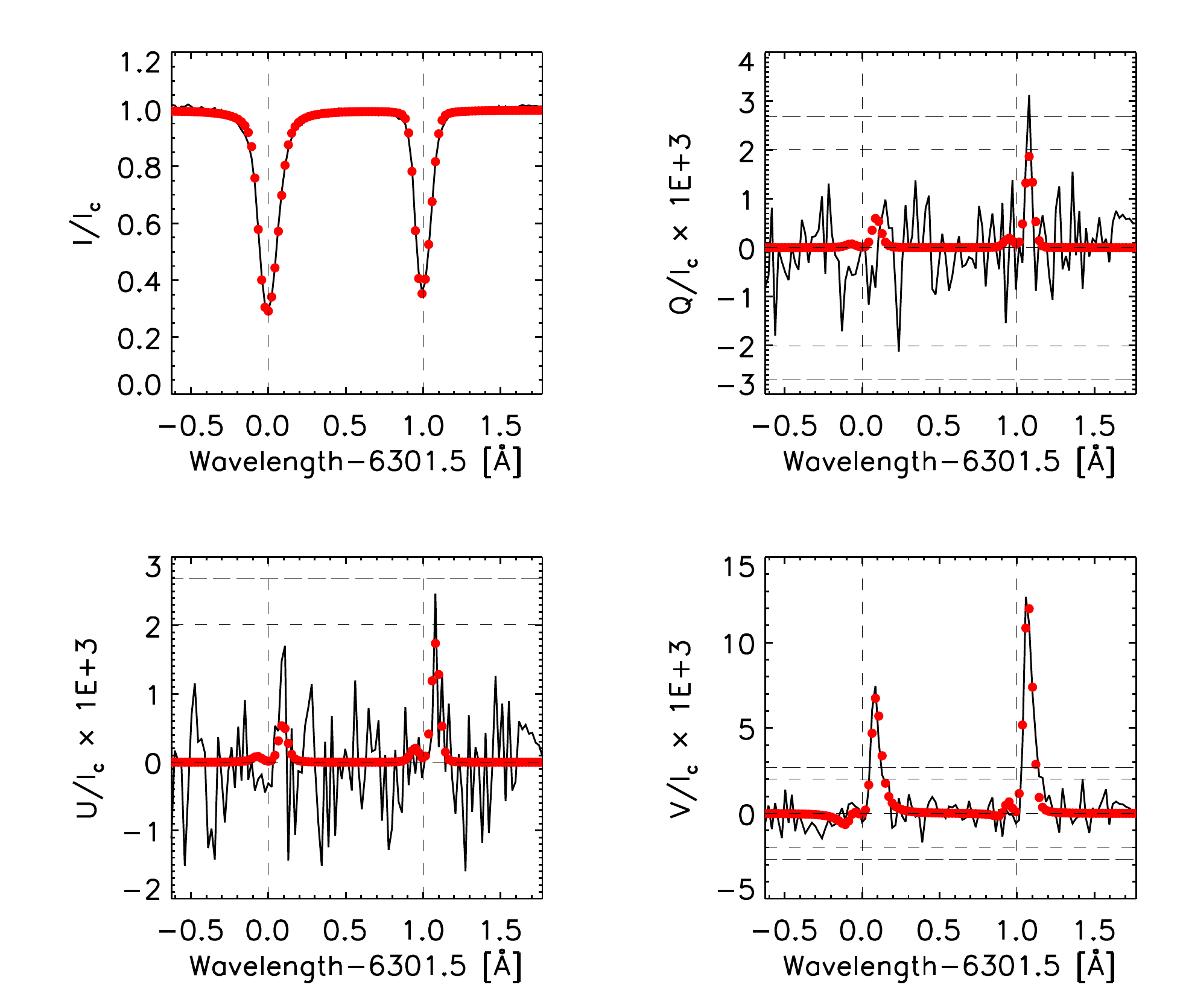}
       {Stokes profiles emerging for the locations marked with the yellow 
       crosses in the left panel of Figure~\ref{comparaI}. The horizontal lines display the thresholds 
       used by the code to identify the \sls s, i.e.,  $\pm3\sigma$ and 
       $\pm4\sigma$. The four left panels (clockwise starting from top left: 
       Stokes I, Q, V, and U) correspond to a case of \bop\ and the four right 
       panels correspond to a case of \rop. 
       The dots represent the best fit profiles found by the inversion
       code SIRJUMP (see section~\ref{sec_inv}).\label{ejestks}}

Let us concentrate on a small region ($21\arcsec \times 16.5\arcsec$) of 
one large scan for showing the location of the blue-only and \rop s. 
The measurements displayed in Figure~\ref{comparaI}, were taken 
on September 24, 2007 at 20 h 42 min UT. The 
spatial sampling is $0.15\arcsec$ and $0.16\arcsec$ in X and Y direction 
respectively; the exposure time is 12.8 s and its location is the center of the solar disk 
($\mu \sim 1.0$). Figure~\ref{comparaI} shows that the blue-only profiles tend to be 
located in the outer part of the granules (blue contours). In contrast, the red only 
profiles are mainly located in the intergranular area (red contours).

\dosfig{width=7cm}{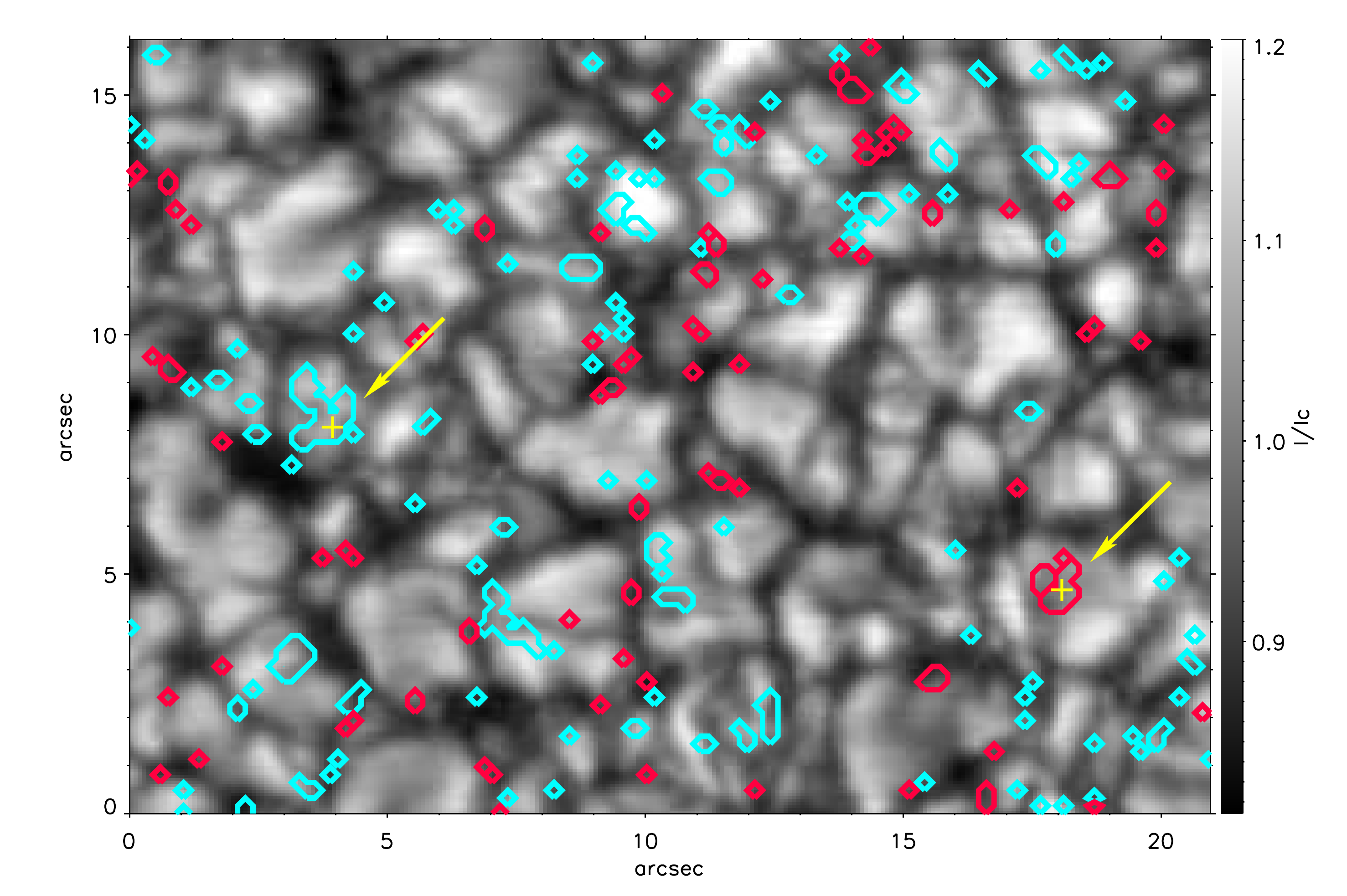}
       {width=7cm}{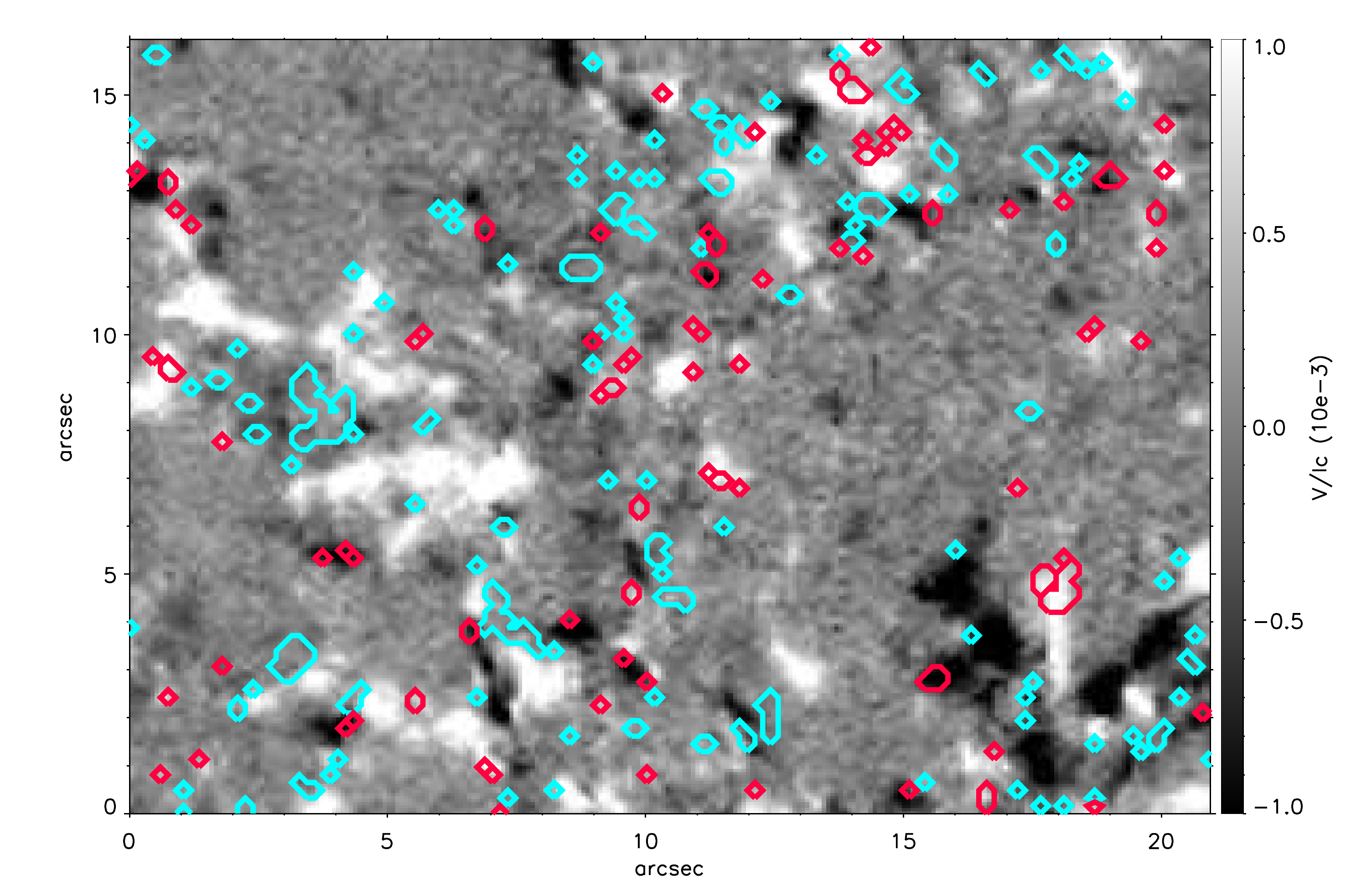}
       {Left: Stokes I map shown in grey scale (part of the observation taken on 
       September, 24 2007) with the blue-only  and \rop\ locations 
       indicated with the respective contours. The yellow crosses pointed by 
       yellow arrows mark the positions of the pixels whose Stokes V profiles 
       are shown in Figure~\ref{perfazules} and~\ref{perfrojos}. 
       Right: Stokes V map. \label{comparaI}}

Figures~\ref{perfazules}  and~\ref{perfrojos} show the spatial distribution 
of the Stokes V profiles in an area of 9 px $\times$ 9 px 
(1.35\arcsec $\times$ 1.44\arcsec) around the yellow crosses 
displayed in Figure~\ref{comparaI}. For reference, the border of 
granules are represented by the black intensity contour.
Figure~\ref{perfazules} makes it clear that there are no regular Stokes V profiles
of opposite polarities close to the blue-only profiles. 
This suggests that the processes leading to the absence of one lobe 
in Stokes V are not related to the horizontal mixing of atmospheres 
with different physical conditions. 
In the case of red-only profiles (Figure~\ref{perfrojos}), there are some 
asymmetric Stokes V signals of opposite polarity nearby, but they seem 
to be related with the expansion of the granular edge in the intergranular lane
(see Section~\ref{sec_temp}). 

\unafig{width=15cm}{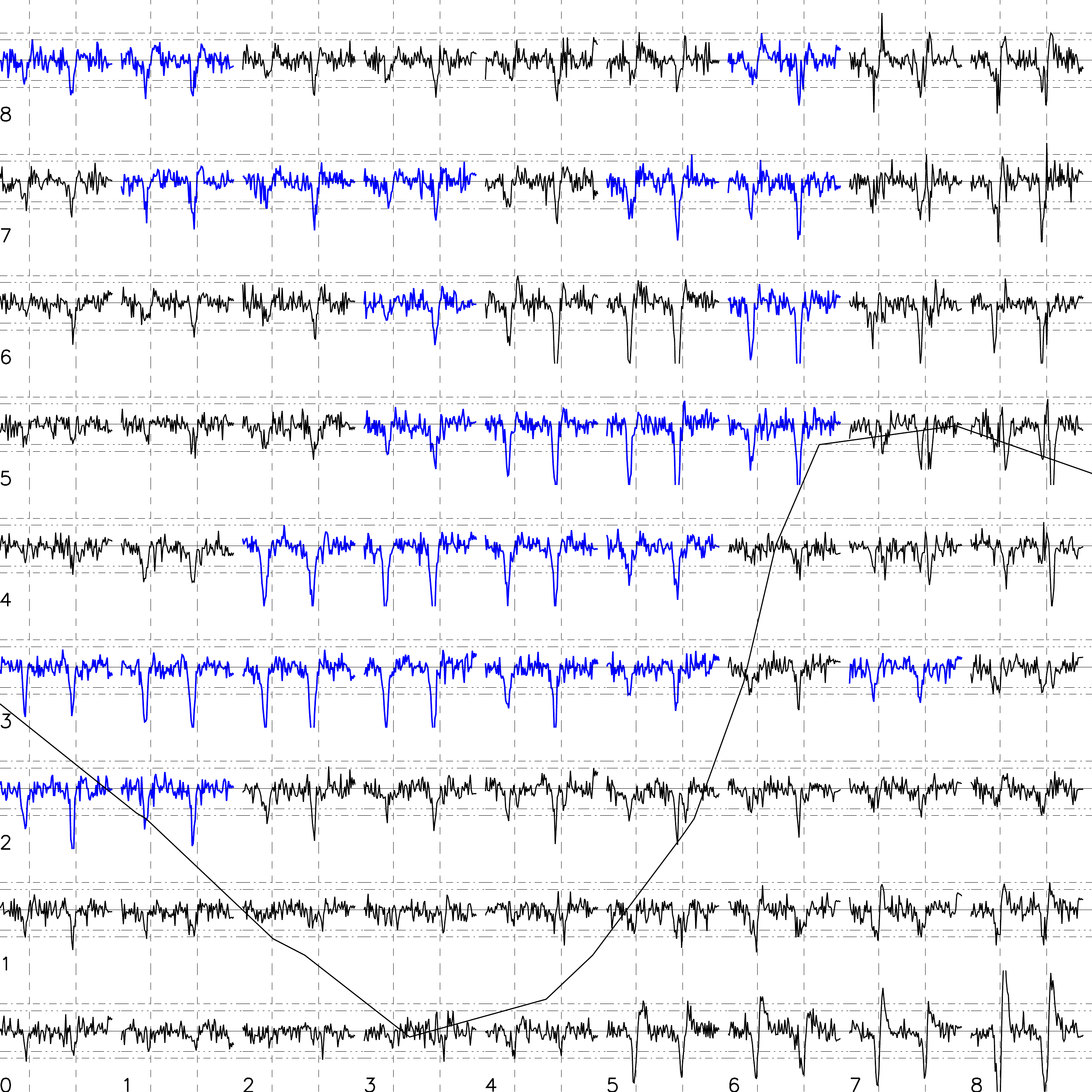}
       {Stokes V profiles shown for each pixel around the yellow cross 
       located at the position (4,8) of Figure~\ref{comparaI}. The horizontal 
       lines mark $\pm 3\sigma$ (dot-dot-dot-dashed lines) and 
       $\pm 4\sigma$ (dot-dashed lines) levels. The vertical dashed lines 
       indicate the nominal position of the core of the 
       \ion{Fe}{1} 6301 \AA\ and  6302 \AA\ lines. The contours indicate the 
       visible edge of the granule. This region 
       corresponds to an area with \bop s. \label{perfazules}}
 
\unafig{width=15cm}{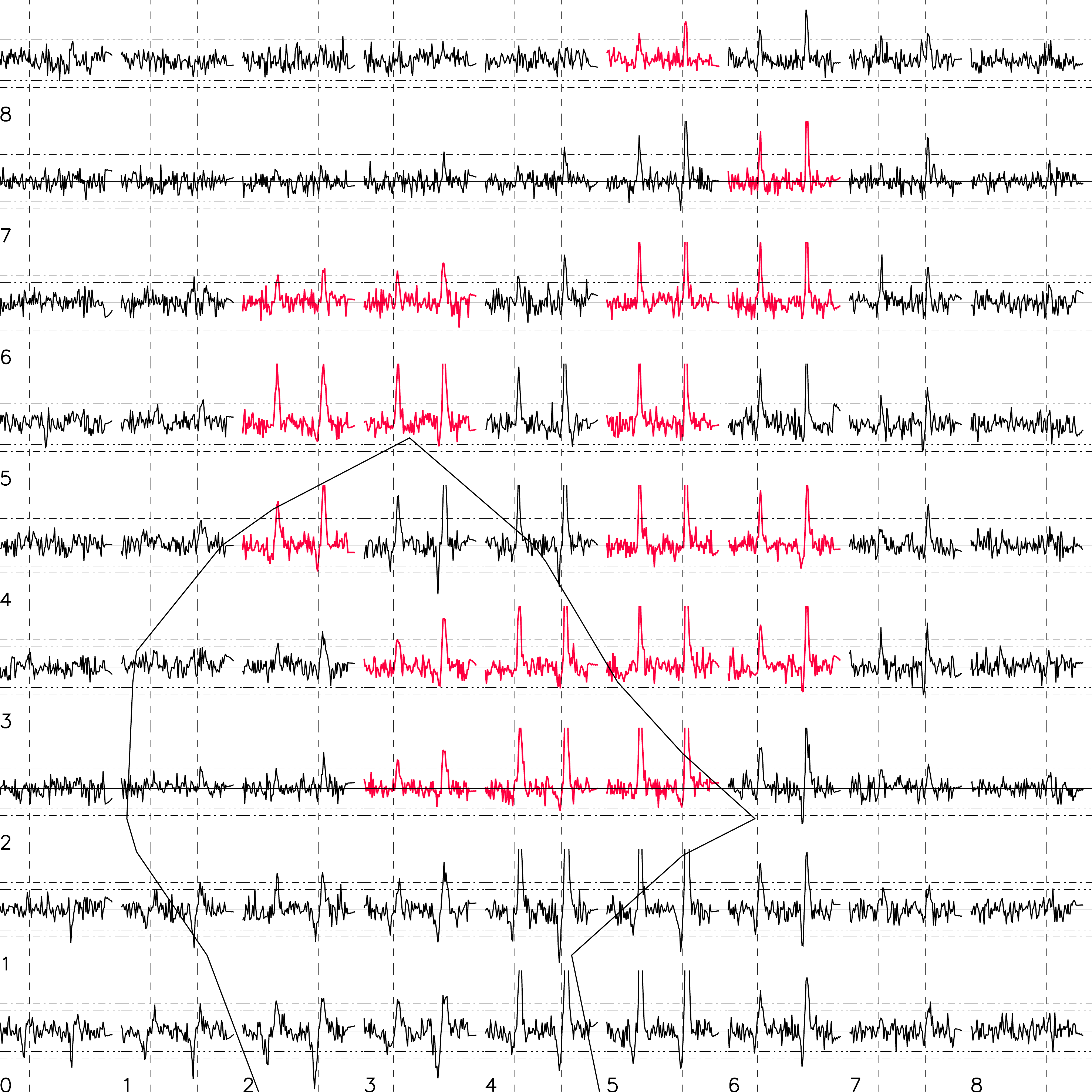}
       {Same as Figure~\ref{perfazules}, for the Stokes V profiles 
       around the yellow cross located at (18,5) in 
       Figure~\ref{comparaI}. This region corresponds to an area with \rop s. 
       \label{perfrojos}}

\section{Physical characterization of their locations \label{propfis}}\label{sec_physical}
We have examined the location of the single-lobed profiles 
on the Stokes I and V maps, dopplergrams, and 
polarization degree maps to understand the general behavior 
with respect to the intensity, velocity and magnetic field. 
In addition, we have calculated histograms for these physical quantities. 
For clarity, we only show the previously selected case, and the 
mean ranges for these quantities using the full data set. 

\begin{figure*} 
  \includegraphics[width=\textwidth]{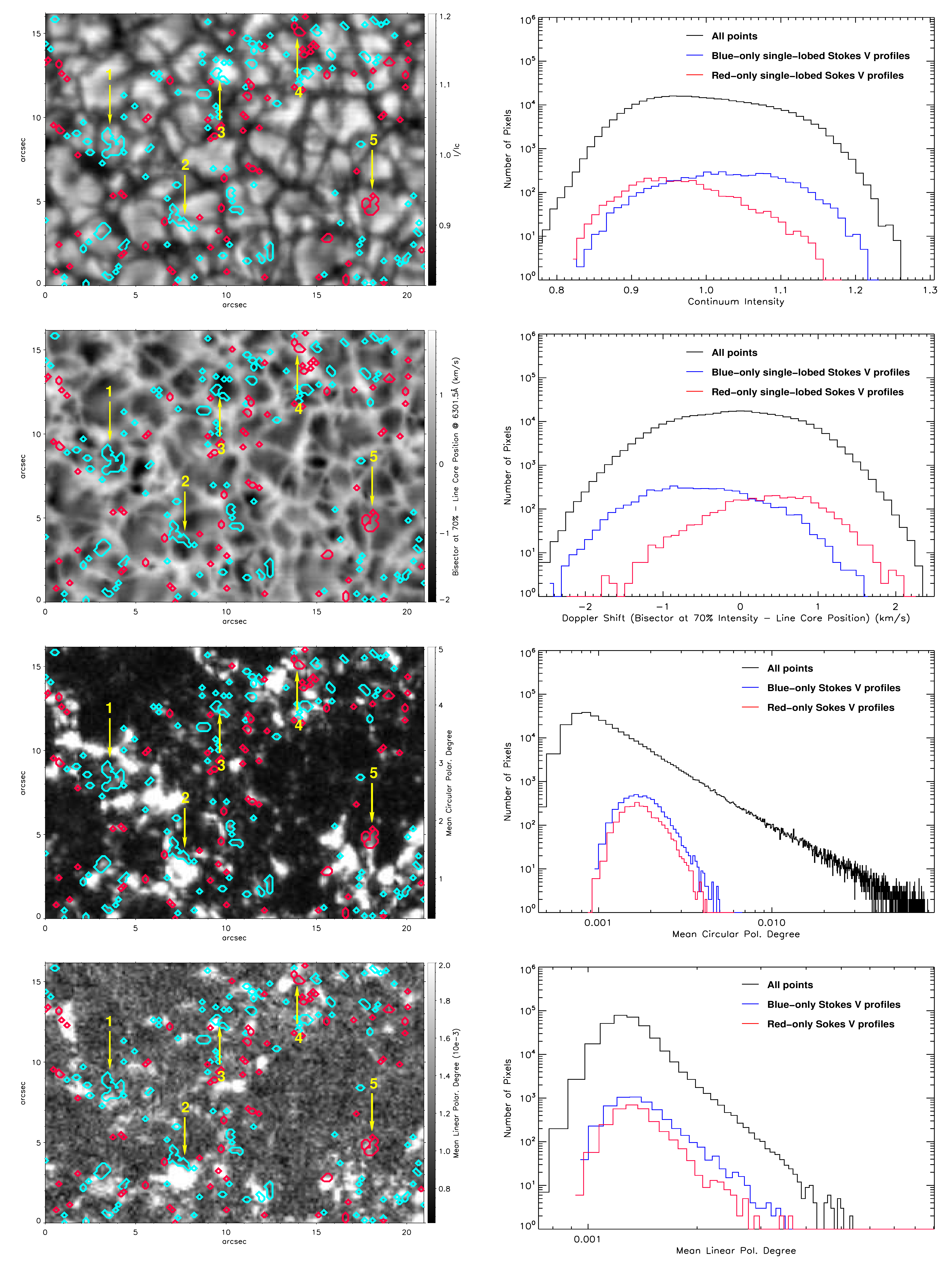}
  \caption{Left column, from top to bottom: spatial distribution 
  of \sls s over-plotted on the intensity, Doppler shift, MCPD 
  and MLPD maps shown in grey scale. The right columns show 
  the corresponding histograms for the whole 
  scan.\label{mapas_histos}}
\end{figure*}

The left panel of the first row in Figure~\ref{mapas_histos} shows 
the Stokes I map of the studied case. We have selected 5 
interesting locations marked with numbers and arrows.
At first glance, this figure tells us two interesting properties 
of the \sls s: (i) The single-lobed profiles tend to appear in small patches. The
red patches are very small and sometimes extend over one pixel only. 
By contrast, the blue patches are well-defined spatial structures, although 
they may look point-like too. (ii) The majority of red-only profiles are located 
in intergranular lanes. The blue-only profiles are very often sited on 
the outer part of the granules, sometimes in intergranular lanes, 
and in a few cases in the center of granules. 

The right panel in the first row of Figure~\ref{mapas_histos} 
displays the histogram of the intensity normalized to the quiet sun continuum. 
Although the distribution is rather broad -notice that the Y 
axis is displayed in logarithmic scale-, it is clear that the 
\bop s are more correlated with the brighter structures than the \rop s.
Except for the scans located very close to the limb, all the 
other data sets show this behavior. On average, 
the distribution associated with the \bop s has the maximum 
at $1.05\pm0.05 I_{c}$ and spans the range 
$[0.85,1.20]\pm0.05 I_{c}$. For the \rop s the maximum 
is located at $0.95 I_{c}\pm0.05 I_{c}$ 
and the range is 
$[0.85,1.10]\pm0.05 I_{c}$. The ranges are very similar 
in both cases, but the distribution of the
 \rop s is always located to the left (darker) 
in comparison to the \bop s,  as it is shown in 
Figure~\ref{mapas_histos}. 
Near the limb, the ranges in intensity are similar for red-only and blue-only
profiles, and the maximum are for both cases located at $1.00\pm0.05 I_{c}$.
Therefore, it is hard to distinguish where these profiles are located. 
This is a consequence of the low intensity contrast at the limb. 
In addition, one should keep 
in mind that observations close to the limb refer to higher layers 
of the solar atmosphere.

The dopplergram and the velocity distribution are shown in 
the second row of Figure~\ref{mapas_histos}. 
Similarly to the intensity distribution, the velocities associated 
with \sls s show a broad range of variation, but there is a trend 
of the \bop s to be associated with upflows (negative velocities). 
The red-only profiles tend to be associated with downflows
(positive velocities). On average, the maximum of the velocity 
distribution for the \bop s is -0.8 km s$^{-1}$, 
within the range [-2,+1] km s$^{-1}$. For the \rop s the 
maximum is 0.8 km s$^{-1}$, with values ranging from 
-1~to +2~km s$^{-1}$. Therefore, their 
velocities are roughly shifted +1 km s$^{-1}$ with 
respect to those of the \bop s. For the \sls s closer to the limb 
the ranges are [-2,+3] km s$^{-1}$ 
and [-2,+2] km s$^{-1}$ for the blue-only and  \rop s, respectively.  
These results are in disagreement with \cite{Sig01}: On average, he 
associated both the blue-only and red-only profiles with downflows, although 
the former had stronger downflows than the latter.

Now we describe the five locations marked in Figure~\ref{mapas_histos} 
that we consider interesting to analyze in more detail. Locations \#1 and 
\#2 are two examples of large size patches. The first extends almost over 
the whole granule, while the second  covers the outer part 
of two granules and the intergranular lanes. Location \#1 sits in an 
upflow region (dark areas in the dopplergram shown in 
Figure~\ref{mapas_histos}); the lower part of location 
\#2 is also in an upflow region, but most part of it is in a transition 
region with velocities close to zero (grey color in the dopplergram). 
Location \#3 is just over the outer part of the granule with a 
high intensity and large upflows. 

In the same figure, the locations \#4 and \#5 show \rop s. 
The former is located in an intergranular lane, with strong 
downflows (bright areas in the dopplergram). Location \#5 is a 
structure nearly $2\arcsec \times 2\arcsec$ in size. It is over
a transition region in the middle of two granules. It is not so 
well-defined as an intergranular region in the intensity map, 
but it is more evident in the dopplergram. In the dopplergram, 
this region is correlated with a narrow bright (downflow) 
intergranular lane and with the grey areas of the outer 
part of the neighboring granules.

Looking at the mean polarization degree maps we find other 
interesting properties of the \sls s. 
The third and the fourth row of Figure~\ref{mapas_histos} 
respectively show the MCPD and MLPD maps (left) and their 
distributions (right). Single-lobed profiles tend to be located 
at the periphery of magnetic flux concentrations showing 
strong linear and/or circular polarization signals.
This fact can be seen at locations \#1, \#2 and \#3, where the 
contours are surrounded by strong magnetic components. 
These locations are close to magnetic structures but not 
exactly on them.  
The locations \#4 and \#5 are partially over magnetic structures 
with weak signals in the polarization degree  maps.
This is the most common behavior for the \sls s, as it can be seen 
in the distributions of the polarization degree shown in 
Figure~\ref{mapas_histos}. The mean circular polarization degree 
histogram follows a linear power law of index $-2.48$ in the 
range [0.0008,0.04]. However, for the
blue-only and \rop s the mean circular polarization degree histogram
follows a sharper linear power law of index $-6.02$ and 
$-6.49$ for blue-only and \rop s, respectively, both in the
range [0.0015,0.005]. The mean linear polarization 
degree histogram follows a linear power law of index $-8.16$ in the 
range [0.0015,0.005]. In a similar manner, the mean linear polarization 
histogram for red-only and \bop s follows a power law of index $-8.15$ 
and $-7.33$ respectively, in the range of [0.0015,0.003]. 
Both the blue-only  and the \rop s are related with modest 
values of the longitudinal component and with the weaker transversal 
field components. In summary, the \sls s are either close to strong 
magnetic structures -but not inside them- or over weak structures 
with both components of the magnetic field.

The locations of the \sls s do not show a preference with the 
polarization sign of structures, i.e., these profiles appear 
with the same probability near positive and negative magnetic fields 
(see Stokes V map in the left panel in Figure~\ref{comparaI}). 

\begin{figure*}
  \includegraphics[width=\textwidth]{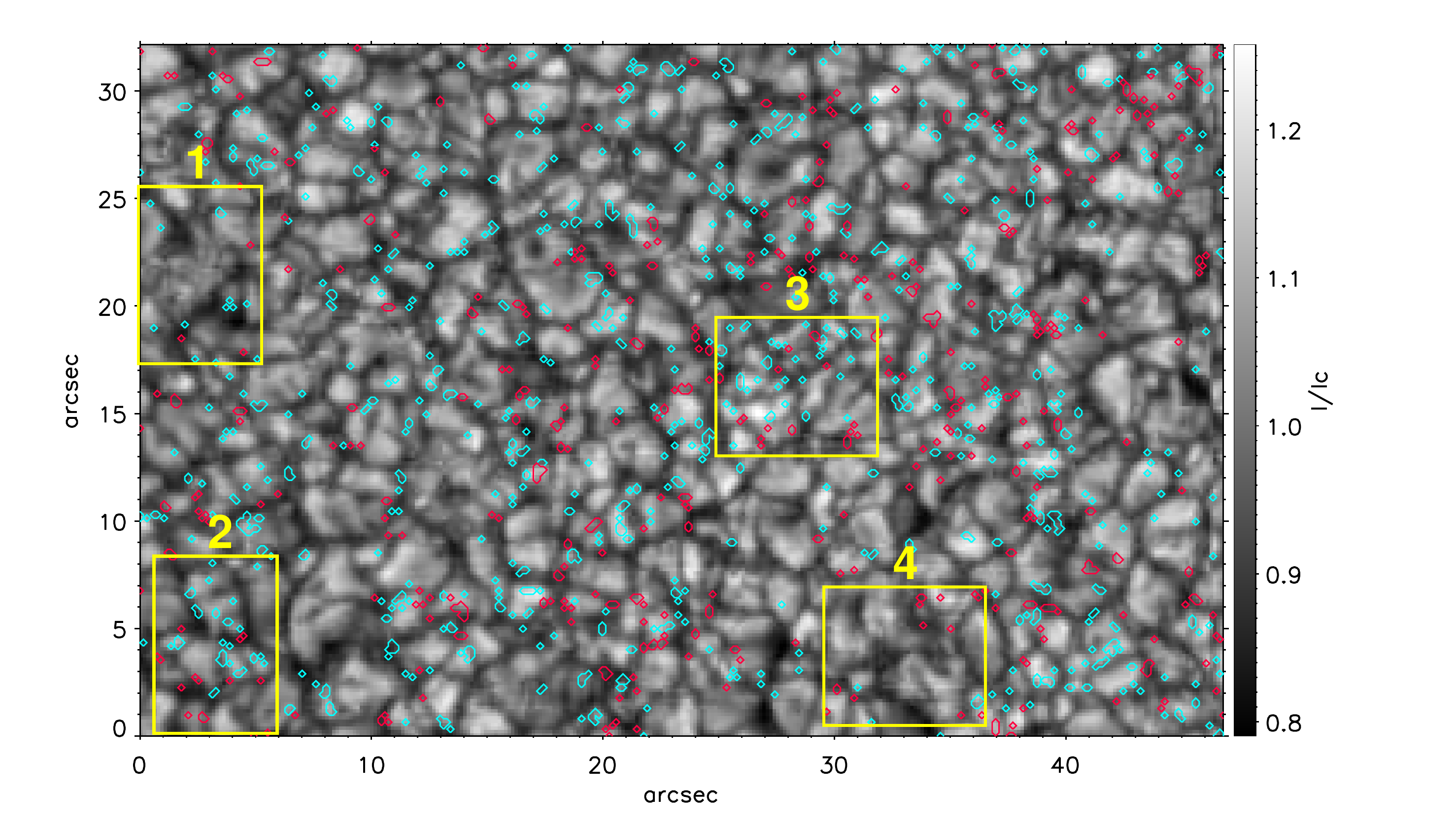}
  \caption{Intensity map showing the spatial distribution of \sls s in quiet sun at the disk center.
       In the selected areas \#1 and \#4 these profiles are located 
       around abnormal granulation patterns because of the presence of the 
       network. The \sls\ can also be seen in normal granulation, 
       as in regions \#2 and \#3.  \label{network0}}
\end{figure*}

If we look in a larger area, as the one shown in Figure~\ref{network0}, 
we can see how the \sls s are around abnormal granulation 
areas (as region \#1 and \#4) but also in normal granulation 
regions (as region \#2 and \#3). This observation 
contradicts the results obtained by \cite{3lobe_Nar10}. 
Indeed, the \sls s are in the borders of the network 
patches -usually showing abnormal granulation pattern 
in continuum images- but never inside them. 
In addition, the \sls s are also present in the internetwork, 
where the granulation has normal properties.

\begin{figure*}
  \includegraphics[width=\textwidth]{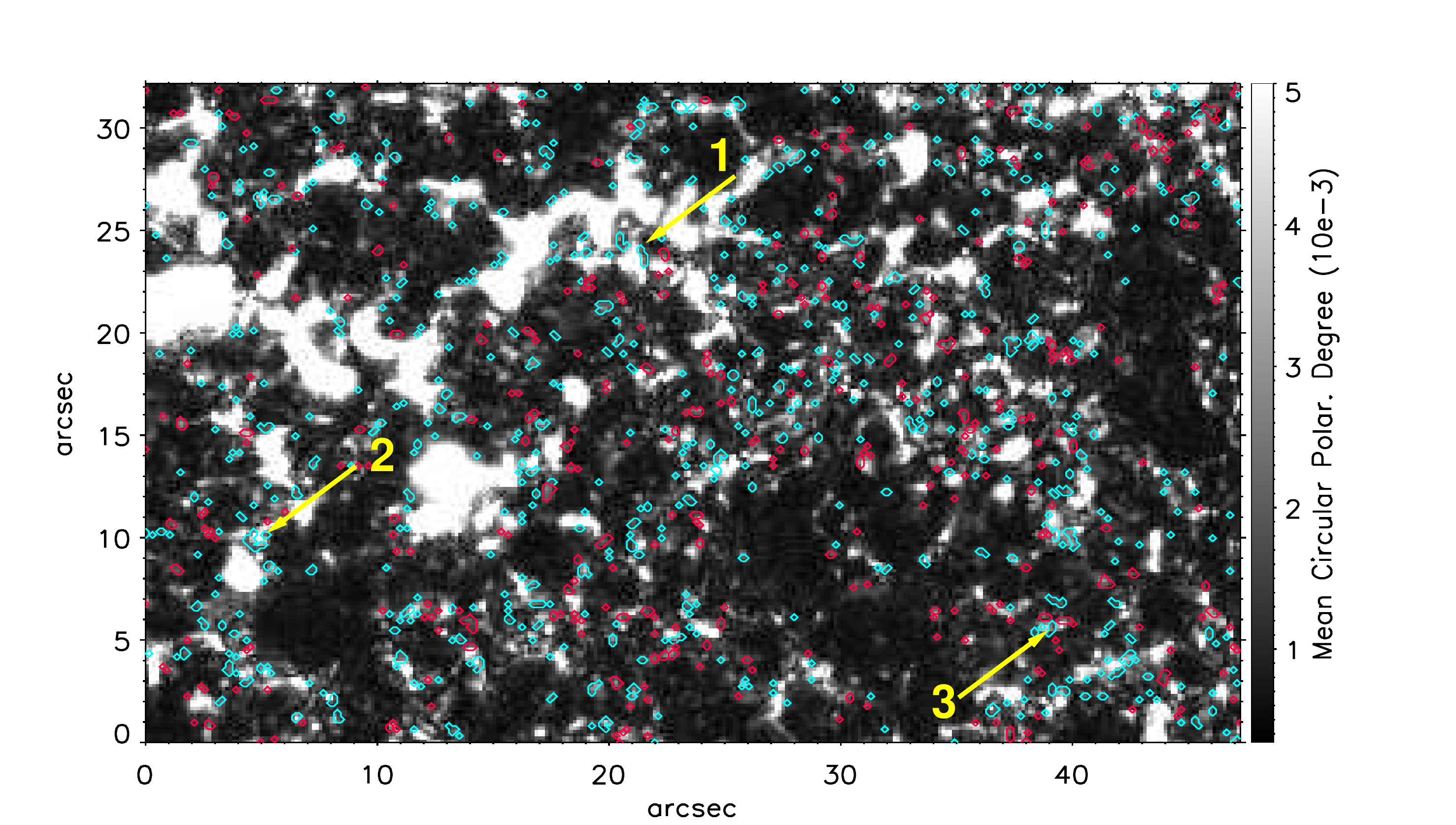}
  \caption{MCPD map of the network and internetwork regions shown in 
       Figure~\ref{network0}, with blue-only and red-only profiles in 
       blue and red, respectively. The arrows mark locations where the 
       blue-only and \rop s are either totally or partially located on structures 
	with moderate strength of the longitudinal polarization signal. 
  	\label{network1}}
\end{figure*}

\begin{figure*}
  \includegraphics[width=\textwidth]{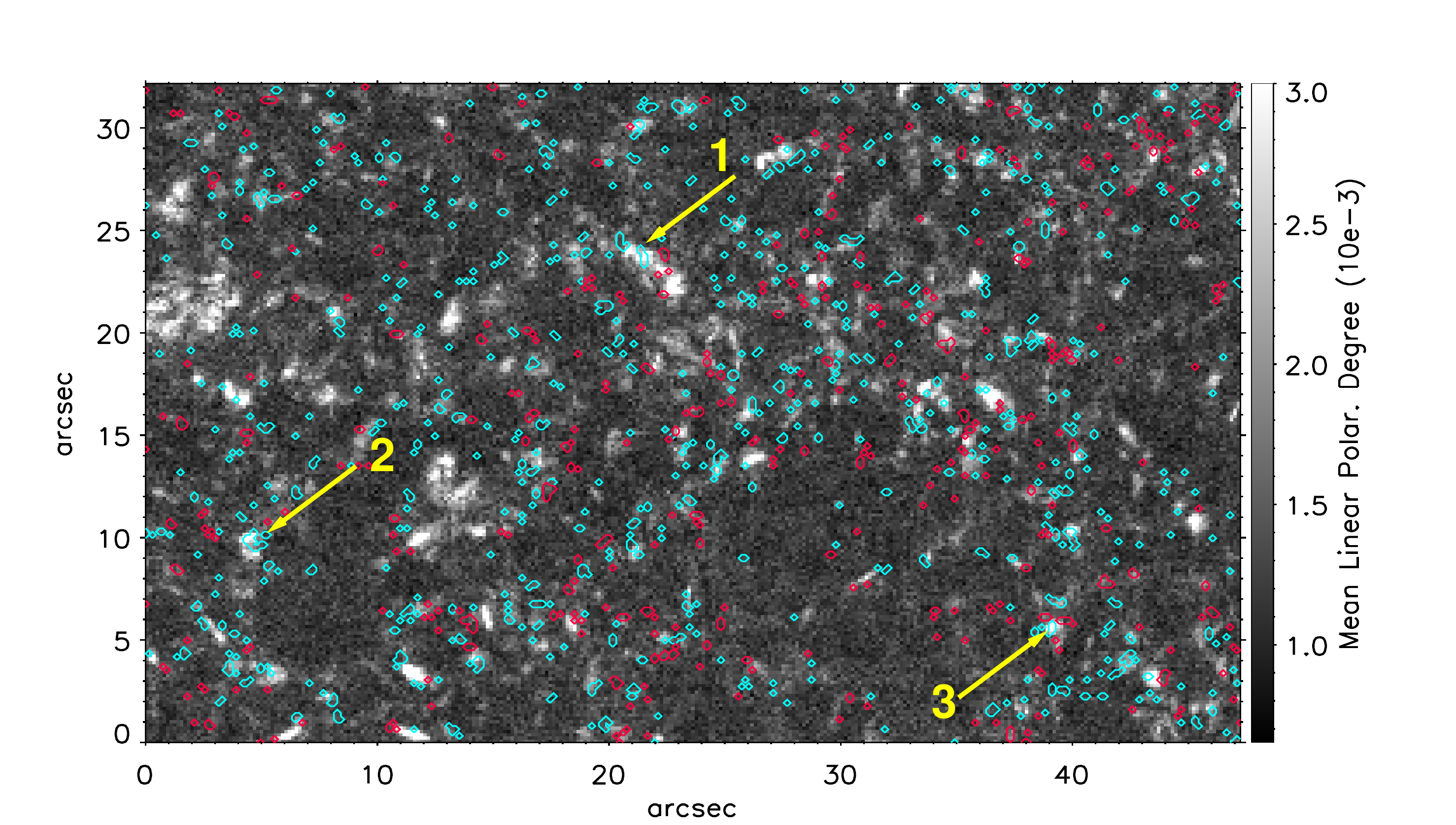}
  \caption{MLPD map (grey scale) of the same area shown in Figure~\ref{network0}, 
       where we over-plotted the blue-only and \rop s in blue and red contours. 
       In this figure we selected several locations (again marked by arrows here) 
       with significant longitudinal polarization signal. These places have also an 
       large linear polarization signal. \label{network2}}
\end{figure*}

Figures \ref{network1} and \ref{network2} show the MCPD 
and MLPD for the same area as in Figure~\ref{network0}. 
The large patches with strong signal form the network. 
The \sls s are never located inside these patches. 
Rather, they surround the network and internetwork elements. 
The arrows mark the locations of several \sls s showing both  
circular and linear polarization signals, i.e., where the magnetic field is 
inclined. Figure \ref{lospeq} shows the MCPD and MLPD maps in an 
area of roughly $5\arcsec \times 5\arcsec$ around the 
selected positions mark with arrows in Figures~\ref{network1} 
and~\ref{network2}. As one can see, location \#1 and \#2 are 
in small patches with linear polarization signal and surrounded 
by network patches with strong circular polarization.
Location \#3 corresponds to an internetwork patch with strong 
linear polarization signal surrounded by small patches with 
weak circular polarization. 

\unafig{width=16cm}{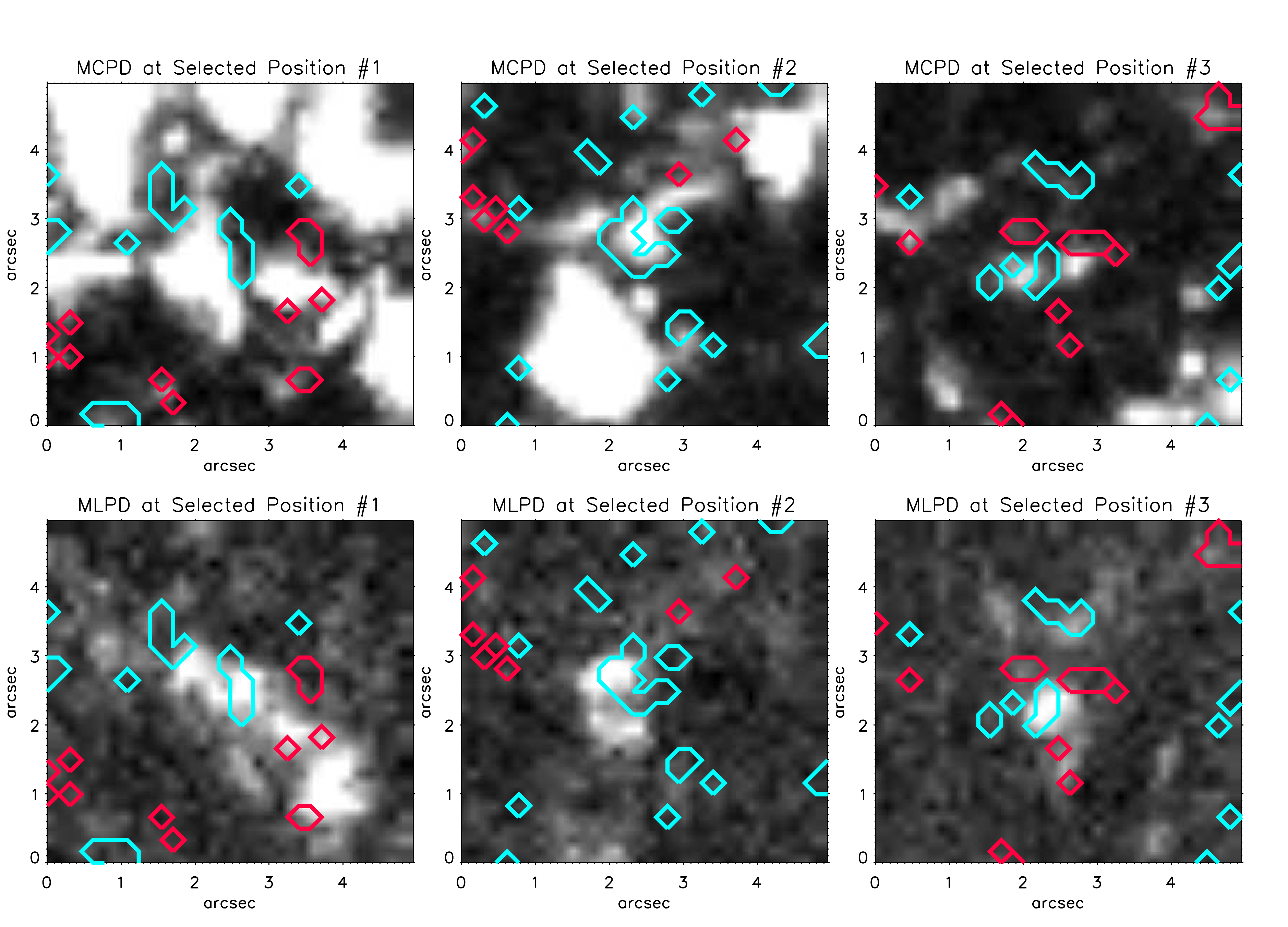}
       {Top: MCPD zoomed maps of the selected positions in Figure~\ref{network1}.
        Bottom: the same zoomed areas for the MLPD displayed in
        Figure~\ref{network2}. The three blue-only patches located in the 
        center of the images have associated transversal polarization 
        signal. The cases \#1 and \#3 are surrounded by structures with 
        longitudinal component of the vector magnetic field. The case \#2 
        has simultaneously both components of {\bf B}. The thresholds for the MCPD and MLPD 
        maps are respectively $5\times10^{-3}$ and $3\times10^{-3}$.\label{lospeq}}  

\section{Center-to-limb variation}\label{sec_solardisk}
As mentioned above, we have examined 72 observing runs to 
search for single-lobed Stokes V profiles. Here, we present the 
behavior of these profiles with respect to their location on the 
solar disk. In Table \ref{latabla} we summarize the statistical 
properties of the profiles. The last 4 columns of the table give the 
following parameters: the signal to noise ratio ({\it S/N}); the 
percentage of \bop s among the Stokes V profiles with absolute signal 
greater than $4\sigma$\footnote{For simplicity, in the displayed panels 
in this paper we use the term {\it `Blue'} as reference to the 
\bop s, and {\it `Red'} as reference of the \rop s. Likewise, 
in Table \ref{latabla} we refer to them as {\it `B'} and {\it `R'} respectively.}; 
the percentage of the \rop s, and the ratio between 
the  blue-only and \rop s (annotated as B/R). 

\begin{figure} 
	\includegraphics[width=9cm]{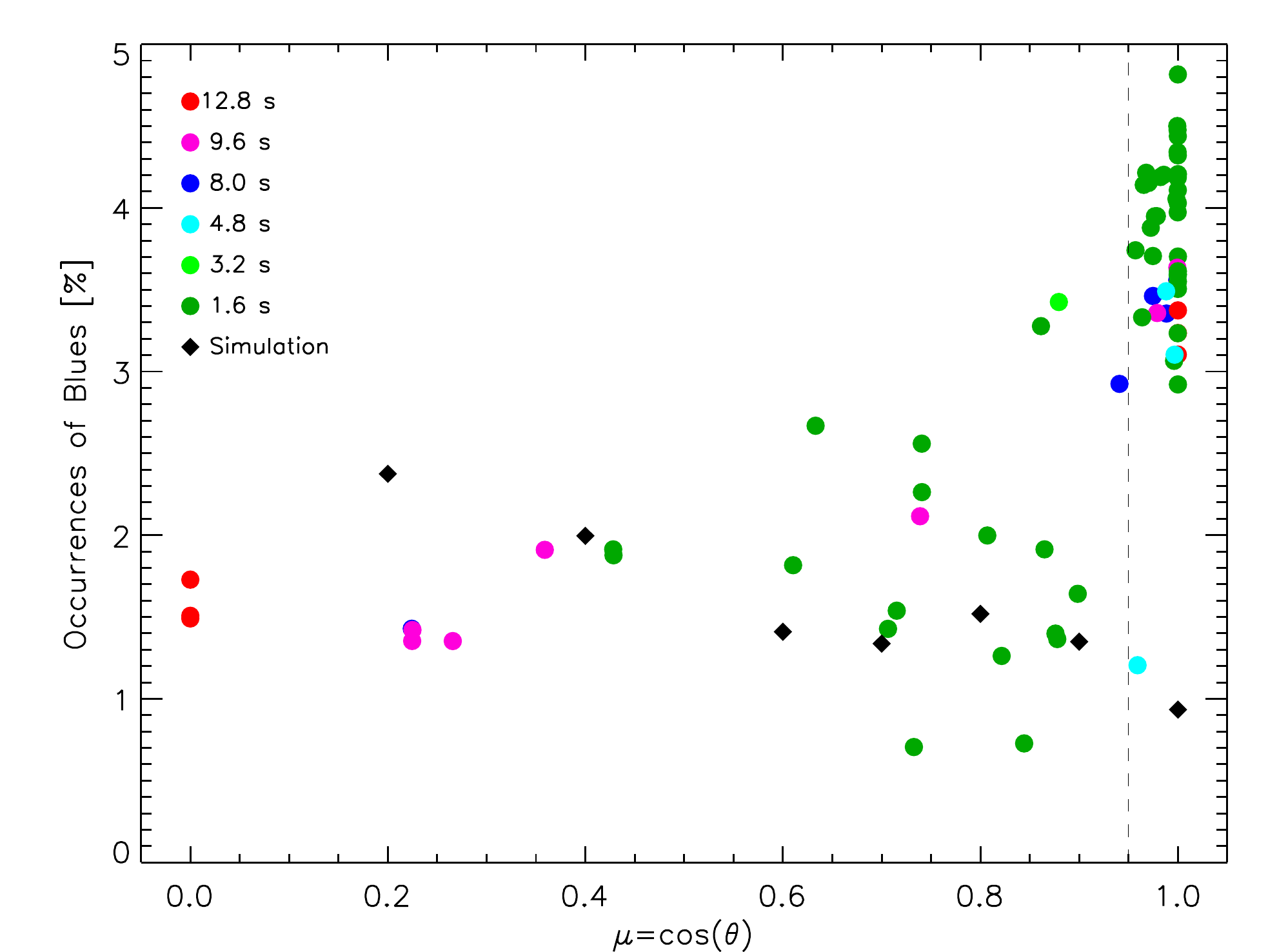}
	\caption{Fractional number of the occurrences of blue-only
        profiles with respect to pixels that have a Stokes-V
        signal greater than $4\sigma$ and their location on the
        solar disk. The vertical dashed line delimits the population I 
	(right) and II (left) (see Appendix). The color code stands the exposure
	time of the observations (see the legend in the plot). Both observations 
	and numerical simulations (diamonds) show a decreasing of \bop s 
	as closer they are to the limb (smaller $\mu$).\label{b_texp}}
\end{figure}

We have plotted the number of occurrences of the blue-only  
and \rop s taking into account their location on the solar disk 
and exposure times in Figure~\ref{b_texp}.
The color of the points codes the exposure time. 
Sometimes, there is overlap between data points. 
For instance, at (1.0,1.6) there is a red point (12.8 s) that 
overlaps a pink point (9.6 s). The fraction of \bop s
in the disk-center is roughly 4\%. The plot shows that the 
fraction of \bop s decreases as the observation 
is made closer to the limb (roughly 1.6\% near the limb). There 
is a decreasing trend in the fraction for the points 
of the same color as they are located closer to the limb (left side of the plot). 
This fact does not depend on the exposure time, i.e., the number 
of \bop\ decreases in the full range of circular polarization signal, not only in
the stronger ones (observed in the center of the solar disk with 
short exposure time), but also in the weaker ones (observed in the limb 
with longer exposure time).  
  
In contrast, the fraction of red-only profiles remains fairly constant for 
different positions on the solar disk (roughly 2\%). There is a slight increase 
of the green points, i.e., from 1.3\% in the disk-center to 3.2\% 
at $\mu \approx 0.46$ (corresponding to observations with exposure 
time of 1.6 s). If we pay attention to the other colored points, the overall 
impression is that the variation with the heliocentric distance 
of the number of occurrence is nearly constant. 

The ratio of blue-only to red-only profiles is shown in Figure~\ref{br_texp}. 
This quantity drops from about 2-4 near the disk center to 0.5 
as the limb is approached. 
This strong variation is mainly due to the strong reduction of the 
number of blue-only profiles toward the limb, not to changes in the red-only 
profiles which remain more or less constant. 

We have calculated the area occupied by the \sls s as the number of these profiles
with respect the total number in every single map. In the region close to the center of
the solar disk, say $\mu < 0.95$, the \bop s occupy 1\% and the \rop s 0.4\% of the 
solar surface. These values change as shown in Figure~\ref{br_texp}, where we see 
that the area occupied by them is 0.9\% and 0.7\% at $0.95<\mu<0.5$, and 
0.5\% and 0.8\% at $\mu<0.5$  for the blue-only and \rop s respectively. Therefore,
on average, the \sls s occupy less than 2\% of the solar surface.

\begin{figure} 
	\includegraphics[width=9cm]{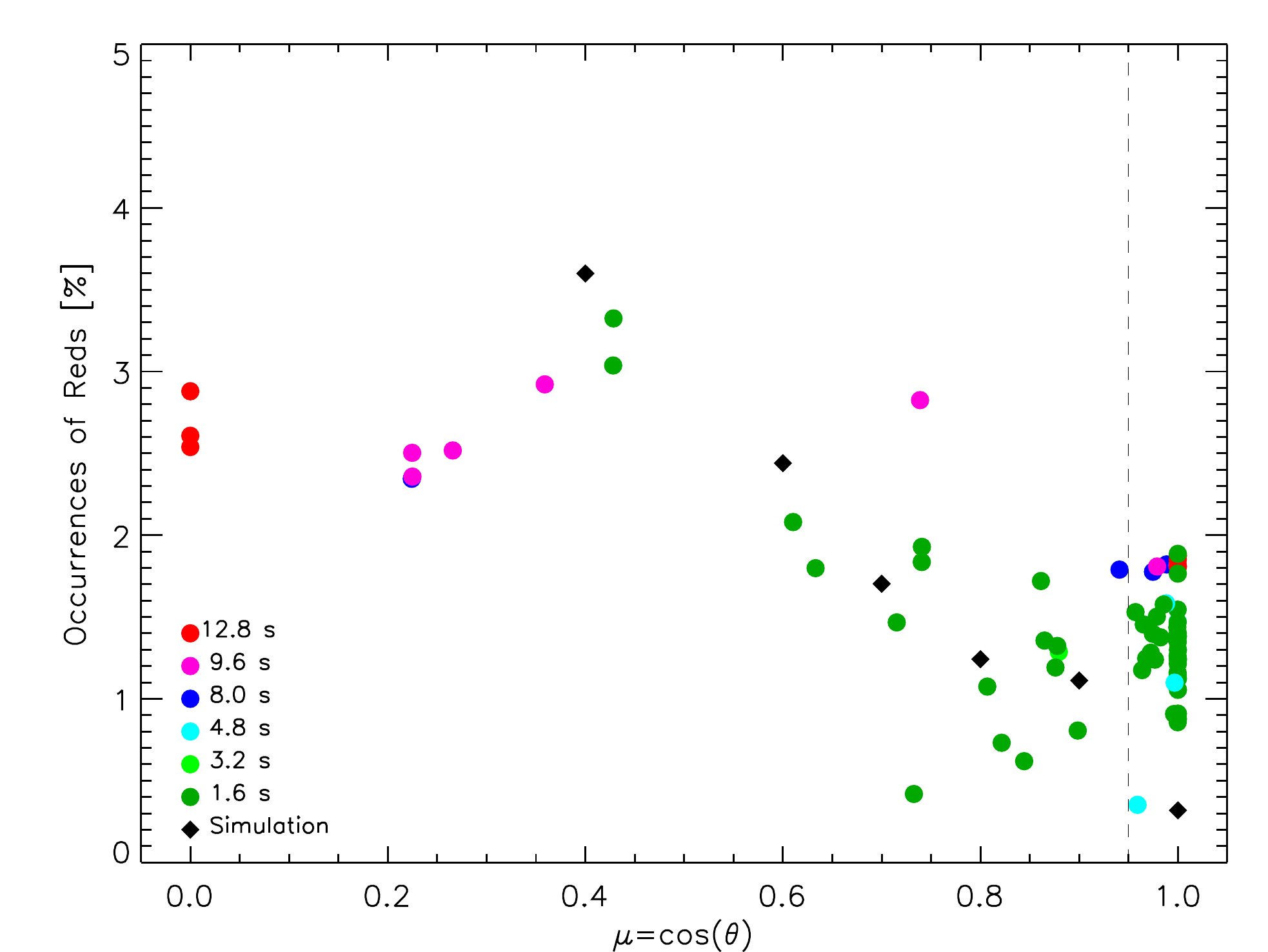}
	\caption{Number of the occurrences of \rop s, following the same 
	symbols as Figure~\ref{b_texp}. \label{r_texp}}
	\end{figure}

\begin{figure} 
	\includegraphics[width=9cm]{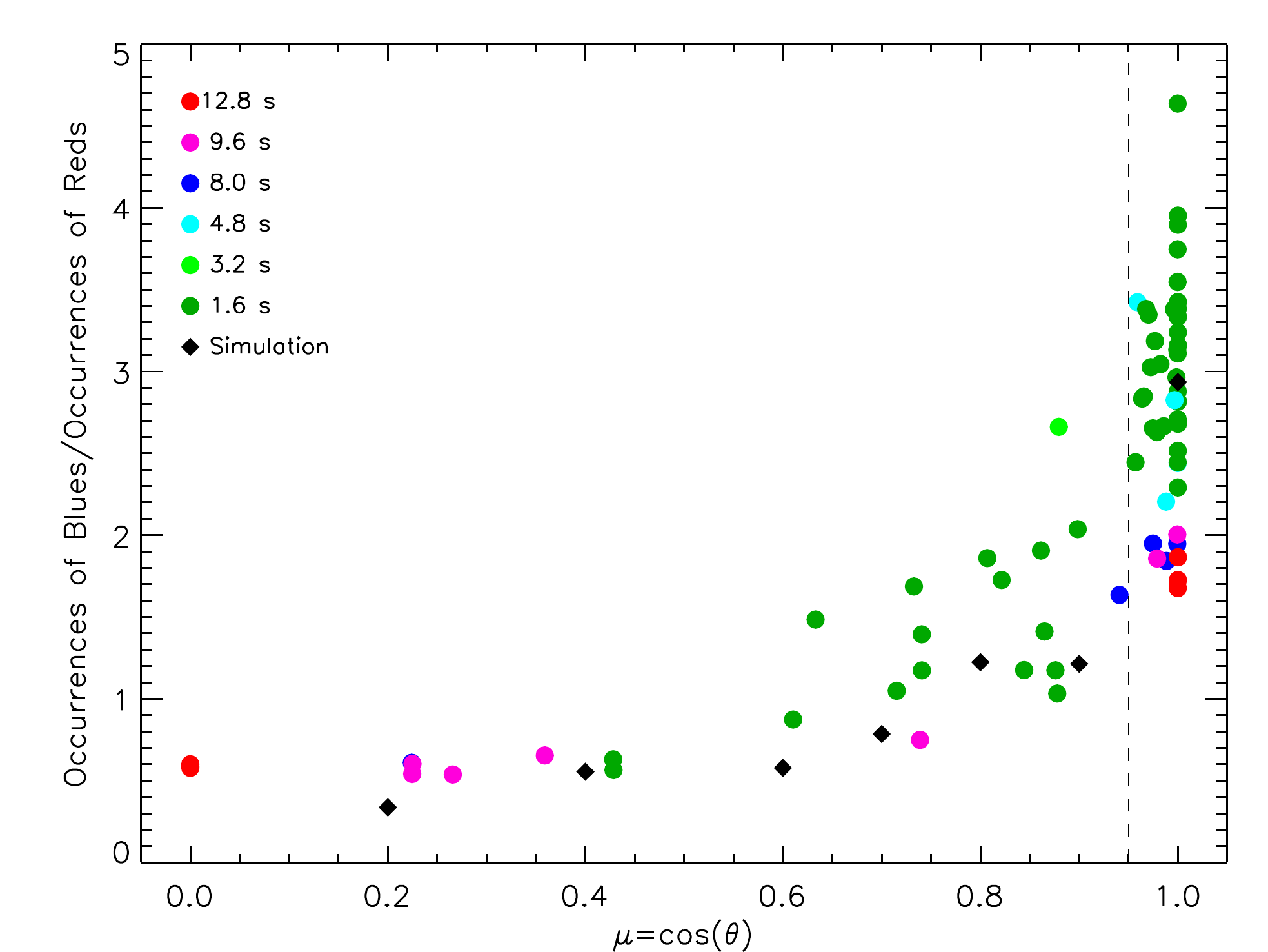}
	\caption{Ratio between the number of occurrences of blue-only 
	and and red-only profiles as a function of the heliocentric angle. 
	The symbols are the same as in Figures. \ref{b_texp} and \ref{r_texp}. 
	\label{br_texp}} 
\end{figure}

\section{Temporal Evolution}\label{sec_temp}


{\begin{figure} 
\includegraphics[bb = 25 0 719 456,  width=0.5\textwidth]{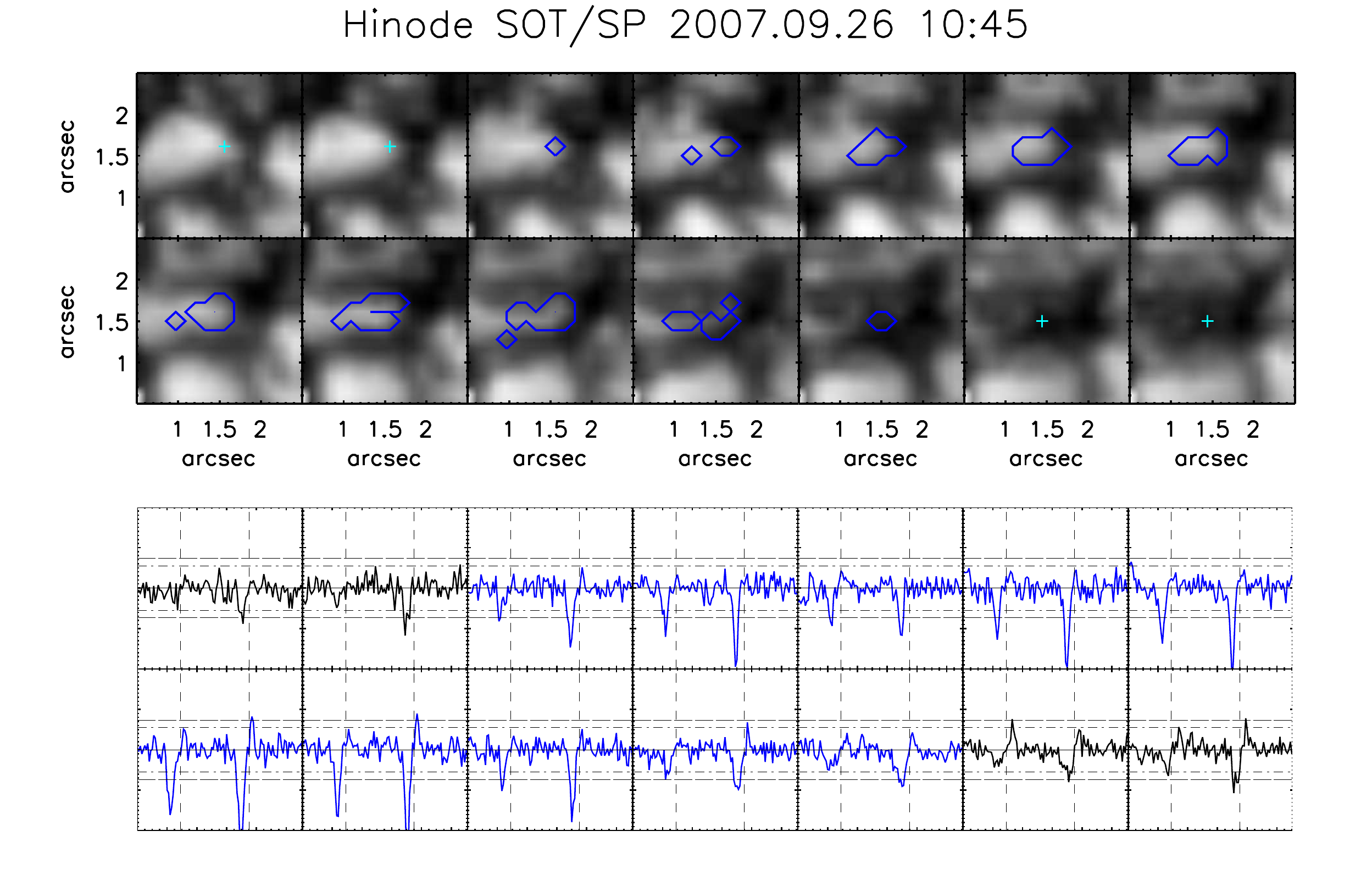}
\caption{Top: Example of temporal evolution of \bop s. (from left to right and from top to
       bottom). Bottom: The most intense \bop s is displayed in blue, while 
       the black line profiles correspond to the location marked with 
       blue light cross before the appearance and after the 
       disappearance of the blue patch. The dashed horizontal lines 
       represent $\pm5\times 10^{-3} I_c$ and $\pm7\times 10^{-3} I_c$ 
       respectively (i.e.,  $\pm3\sigma$ and $\pm4\sigma$). The vertical 
       dashed lines mark the central position of \ion{Fe}{1} 6301 
       \AA\ and 6302 \AA.\label{evol_azu_ii}}
\end{figure}}

We have chosen a raster scan of a  quiet sun area to show how the patches 
of single-lobed Stokes V profiles evolve in time. The observation was 
taken on September, 26 2007 at 10:45 UT, with a spatial sampling of 
0.15\arcsec\ and 0.16\arcsec\ in X and Y direction, a spectral sampling 
of 21  $m$\AA\ and an exposure time of 1.6 s. The scan covered an area 
about 2.7" wide at a cadence of 35 s. Figure~\ref{evol_azu_ii} shows the 
temporal evolution of a small area during 8.2 minutes (14 consecutive raster scans). 

The temporal evolution of a granule is well centered in the FOV. 
Although we are interested in the evolution of the \bop s that appear in the 
second and the third panels, we show some previous and later instants 
for this small area, too. The blue light crosses mark the positions 
where the \bop s {\it could be} before and after they are observed. The 
third image of the top row of Figure~\ref{evol_azu_ii} shows that the 
blue patch appears at the right part the granule, becoming larger and 
larger in the next images. Then, it splits in smaller patches, but with 
the \bop s keeping always concentrated in the right part of the 
granule. Finally, the blue patch disappears in a large-sized 
intergranular lane. 

The bottom panel of Figure~\ref{evol_azu_ii} 
shows the temporal evolution of the 
observed circular polarization signals.
The profiles displayed in blue are the stronger \bop s in the blue 
patch, while the ones in black  are the profiles corresponding to the 
manually selected positions (blue light crosses in the top panel 
of Figure~\ref{evol_azu_ii}). As in previous plots, the 
horizontal dashed lines mark $\pm 3\sigma$ and $\pm 4\sigma$, 
that roughly correspond to $\pm5\times 10^{-3} I_c$ 
and $\pm7\times 10^{-3} I_c$ respectively. The vertical 
dashed lines mark the center of the \ion{Fe}{1}  6301 \AA\ and 
6302 \AA\ spectral lines. 
The \sls s appear after the granule is formed, and 
the strength of the signal and size of the patch increases with time until the 
granule starts to disappear.
From the available time-series, before of the images 
shown in the plot, the profiles in this area are 
noisy and start to show a single lobe (similar to the first profile
displayed in Figure~\ref{evol_azu_ii}). 
At these instants, note that the Stokes V signal of \ion{Fe}{1} 6301 \AA\
is negligible. 
Later, this weak signal increases until the first proper 
\bop\ appears. At the end, the Stokes V profiles are weak but show 
regular shapes with two opposite sign. These signals persist during 
several minutes after the disappearance of the single-lobed profiles, 
as can be seen in Figure~\ref{evol_azu_ii}. 

To better illustrate this, Figures~\ref{perf_antes} and \ref{perf_despues} 
show the spatial distribution of the Stokes V profiles in the first and 
last images of the sequence respectively (Figure~\ref{evol_azu_ii}). 
At positions (4,4) and (5,4) of Figure~\ref{perf_antes}, we can see very 
tiny \sls s and, in their surroundings, at (2,2) and (3,2), a few asymmetric 
profiles with very small signal, or noisy profiles. In Figure~\ref{perf_despues},
around position (4,4), there are also a few profiles with strong 
asymmetry, although they are not single-lobed profiles. For all the time 
series of this example no linear polarization signal is detected. 
Similar behavior has been observed in other scans in the 
sense of the evolution of the \bop\ amplitude and the size
of the blue-only patch. The disappearance of the \bop\ and 
any signal in Stokes V evolve together with the granule, i.e., both
structures, the granule and the Stokes V signal decrease and collapse.

The evolution shown in Figure~\ref{evol_azu_ii} suggest a failure of 
emergence of a tiny magnetic structure which at the end of its life is 
submerging with the disappearance of the granule.


\unafig{width=16cm}{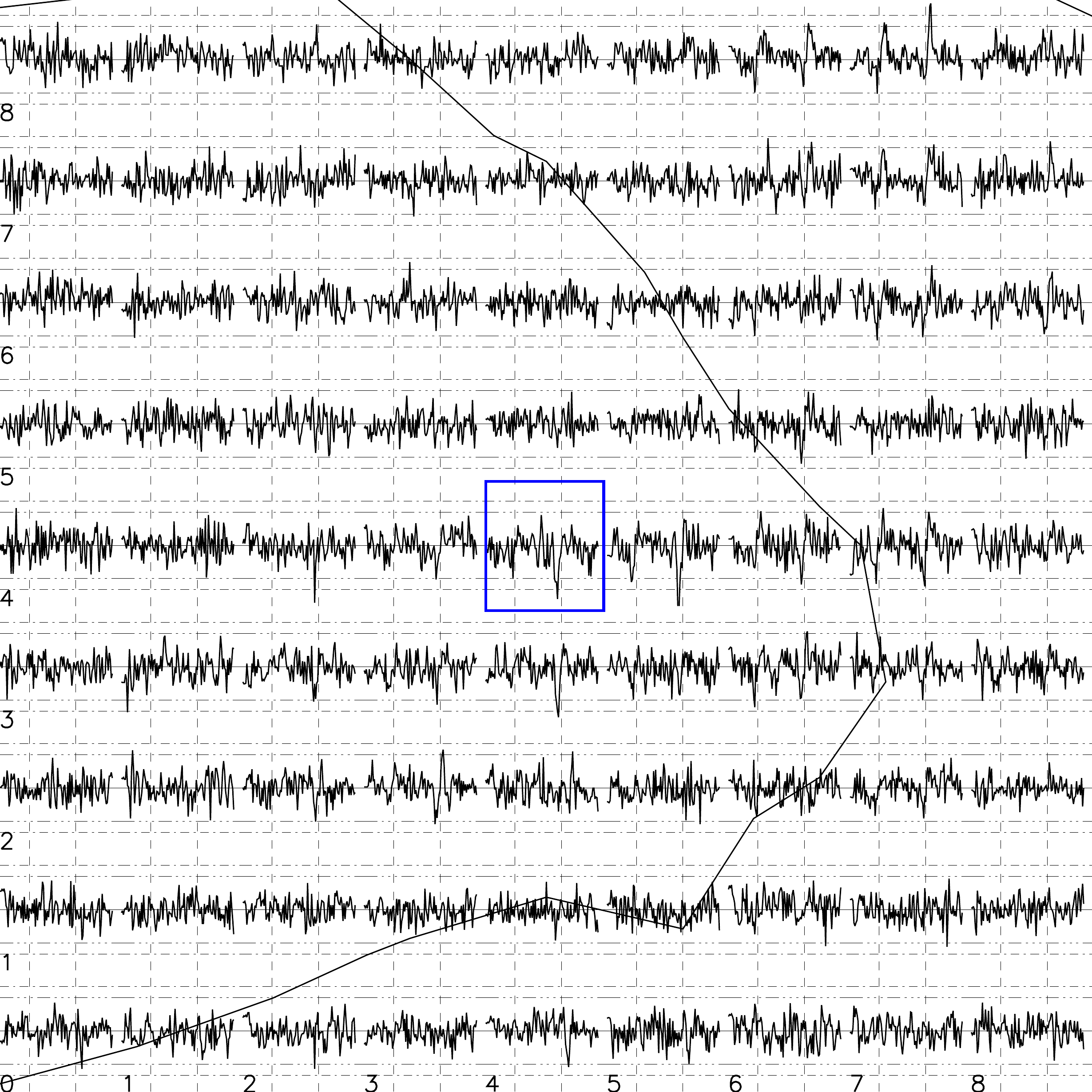}
       {Spatial distribution of the Stokes V profiles around the blue 
       light cross of the first image of the first row of Figure~\ref{evol_azu_ii}. 
       The profile corresponding to this position is enclosed in the blue light frame.
       The dashed lines correspond to $\pm5\times 10^{-3} I_c$ 
       and $\pm7\times 10^{-3} I_c$ (i.e., $\pm3\sigma$ and 
       $\pm4\sigma$). The overlapped black line displays the intensity contour
        of the granule. \label{perf_antes}}

\unafig{width=16cm}{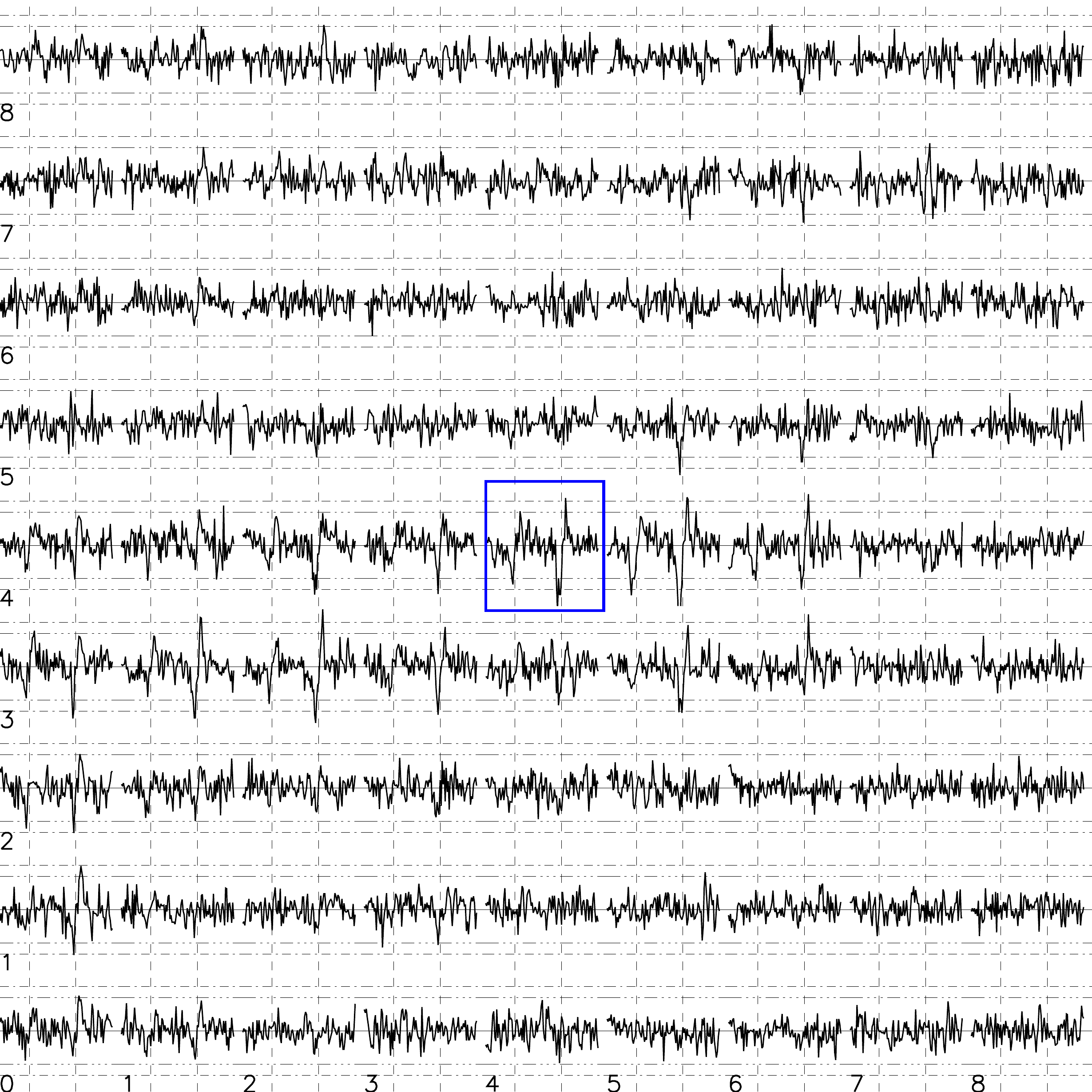}
       {Similar to Figure~\ref{perf_antes}, but for the position marked with a 
       blue light cross in the last image of the second row of Figure~\ref{evol_azu_ii}.
       This position is located in a large intergranuliar lane, far away of the 
       granule border. \label{perf_despues}}

{\begin{figure} 
\includegraphics[bb = 40 0 550 680,  width=0.5\textwidth]{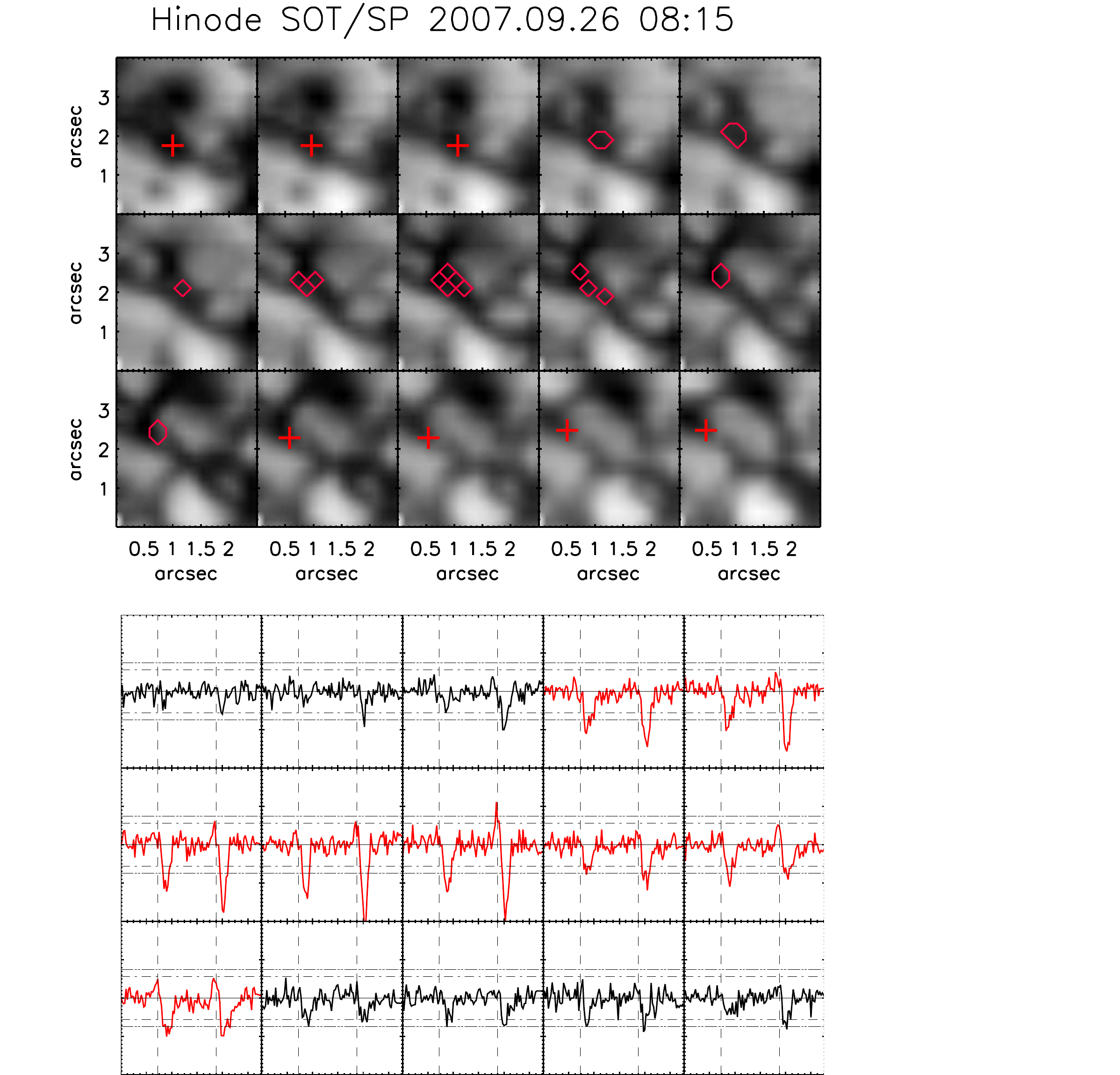}
\caption{Top: example of temporal evolution of \rop s. (from left to right and from top to
       bottom). The lapse between images is 35 s. Three steps before and after of the
       detected \rop s are shown and marked with red crosses.
       Bottom: in black, the Stokes V profiles corresponding to the red crosses. In
       red, the \rop s.
       \label{evol_rojo}}
\end{figure}}

The size of the red-only patches is of the order of the spatial resolution of Hinode, 
and their lifetime (few minutes) is considerable shorter than of the \bop s. 
Figure~\ref{evol_rojo} shows an example of  evolution of \rop s. They were observed
 on September, 26 2007 at 08:15 UT. The red patches are mainly located in the 
intergranular lane, a very changing and dynamic region. The amplitude in Stokes 
V varies in small spatial scales along the intergranular lane. There are three 
steps before and after (detection) of the red patch. At the beginning, 
it is located inside of an irregular-shaped interganular lane, becoming well-defined shape 
as the granules around appear. The patch size seems to be constrained to the 
size of the intergranular lane. In the bottom of Figure~\ref{evol_rojo} is plotted the 
evolution of the profiles. The ones in black correspond to instants before and after the code 
detects the presence of only one lobe. The profile evolution starts with a small \sls s 
- below $\pm3\sigma$ level - increasing the signal as the patch is becoming larger.
The amplitude of the single lobe profile increase with the size of the red patch. 
When it reaches the largest size, it occupies roughly half of the intergranular 
lane (3$^{rd}$ image in the 2$^{nd}$ row of Figure~\ref{evol_rojo}). 
Afterwards, it decreases in area and amplitude until the single lobe 
has small signal and the code does not detect it. For a large period of time, 
there are regular Stokes V profiles next to the weak single lobed profile. 

Figures \ref{distro_rojo_1} and \ref{distro_rojo_2} show the spatial distribution of the
Stokes V profiles around the positions marked by the first and last red cross of
 Figure~\ref{evol_rojo}. The profiles show in Figure~\ref{evol_rojo} are located at 
[4,4] in the two cases. At the beginning of the evolution (see 
Figure~\ref{distro_rojo_1}) there is an area located in the left upper part of the panel 
where the profiles have rather symmetric Stokes V profiles. In this region,
the intensity images (see top panel of Figure~\ref{evol_rojo}) 
show an emergence or 
rise of a small granule (left top to the cross) and expansion of the big granule (left
to the cross). However, in the region where the red patch will appear,  
there are some small \rop s, that becomes \rop s (i.e., detected by the code) three 
images later.  Similarly, at the final stage of the evolution  (see Figure~\ref{distro_rojo_2}), 
there are some symmetric Stokes V profile to the right of the selected profile.
These symmetric profiles become stronger with time as the granule occupy the 
intergranular lane, while the signal of the \rop s (and the profiles located to the left) 
drops below $3\sigma$. The \rop s do not show significant linear polarization signal 
during their evolution. Very weak linear signals can however be seen in the 
surrounding pixels with symmetric Stokes V profiles, as mentioned. 
In summary, the \rop s are well located in the intergranular lane during their life,
even when they are surrounded by symmetric Stokes V profiles. Some \rop s can 
possibly appear in the outer part of granules. In this case they do in very small, 
isolated patches. We have verified that these red isolated patches are related with 
locations where there were bigger red patches previously, 
 and they have eventually been occupied by the expanding outer part of the 
granule.      

\unafig{width=16cm}{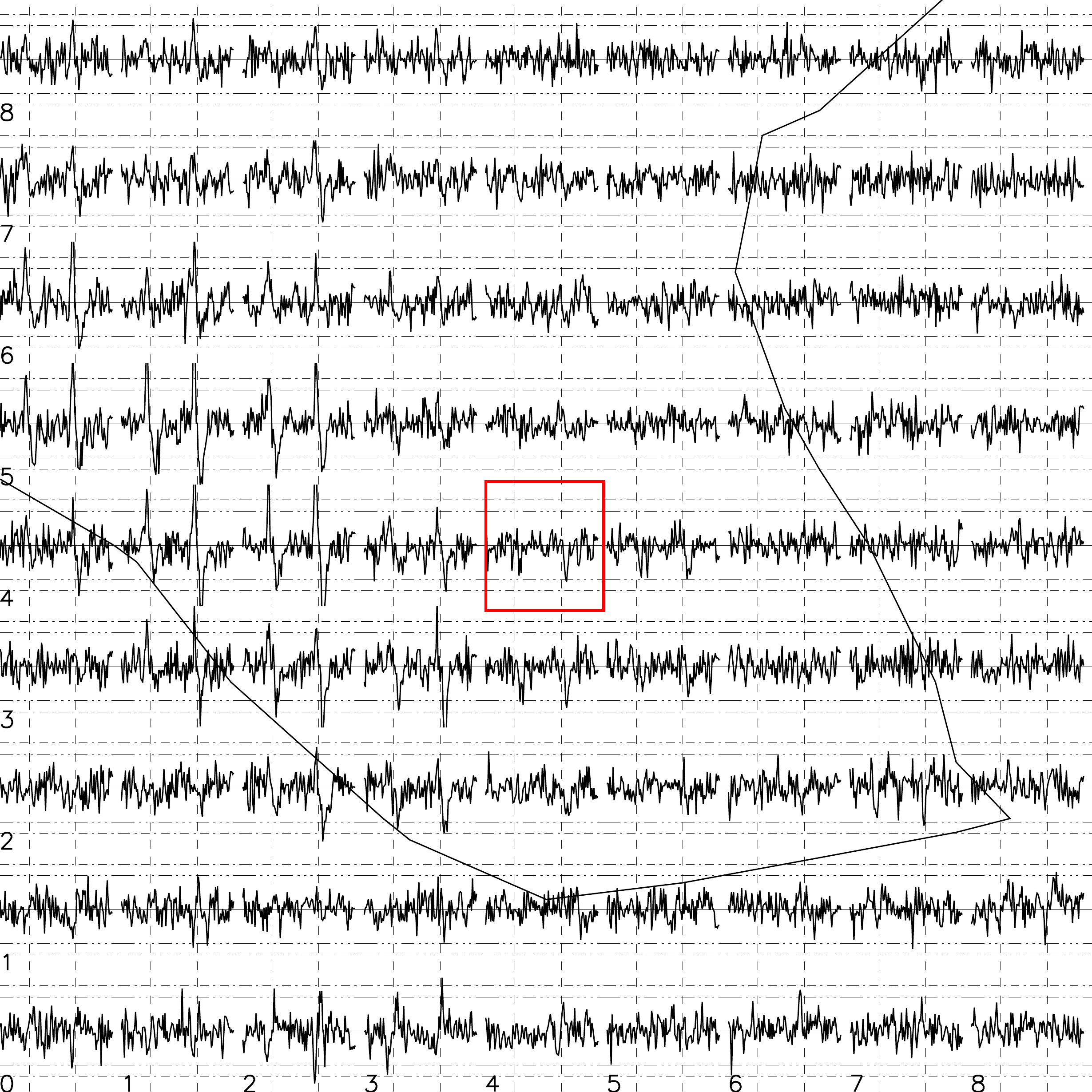}
       {Spatial distribution of the Stokes V profiles around the red  
       cross of the first image of the first row of Figure~\ref{evol_rojo}. 
       The profile corresponding to this position is enclosed in the red frame.
       The dashed lines correspond to $\pm5\times 10^{-3} I_c$ 
       and $\pm7\times 10^{-3} I_c$ (i.e., $\pm3\sigma$ and 
       $\pm4\sigma$). The overlapped black line displays the intensity contour
        of the granule.\label{distro_rojo_1}} 

\unafig{width=16cm}{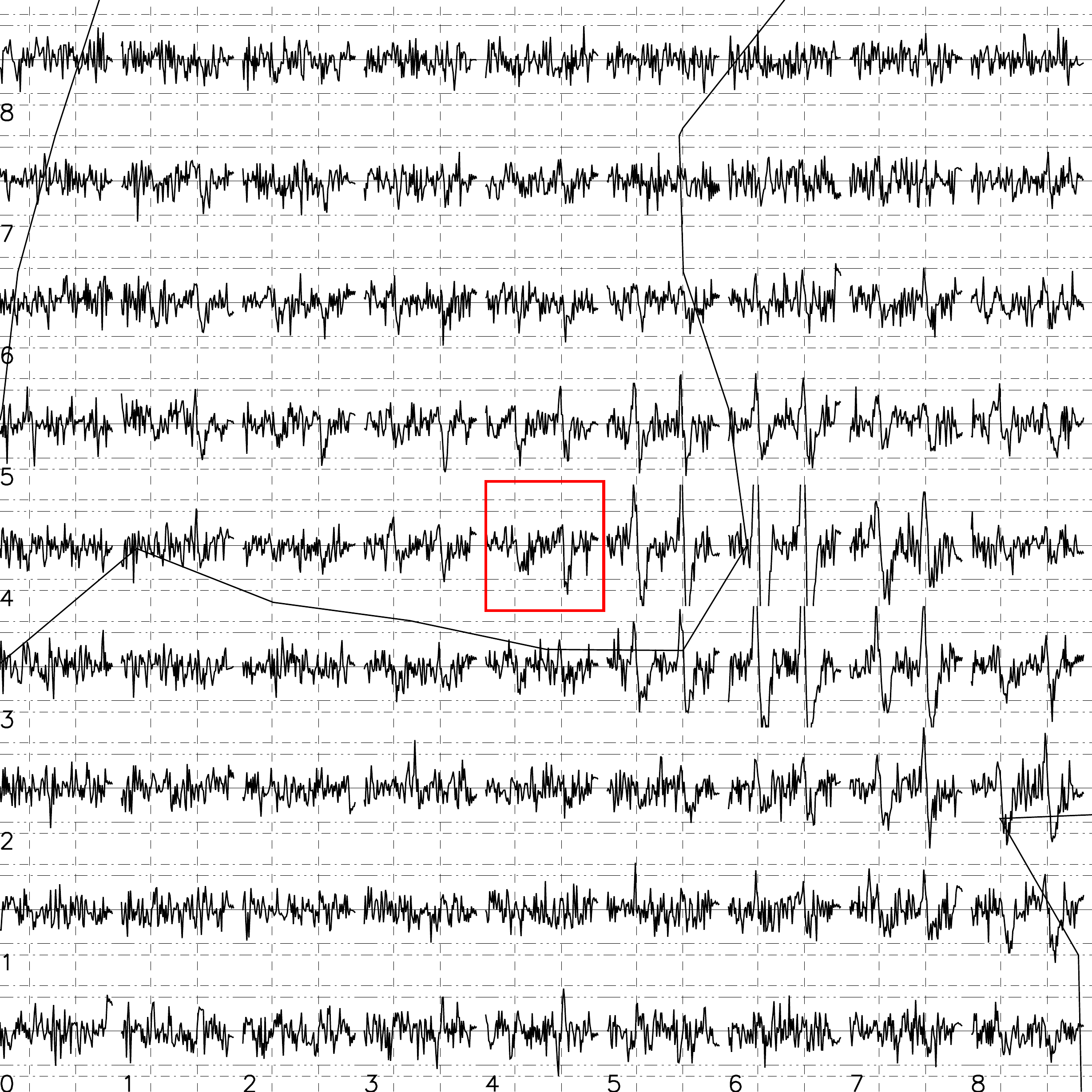}
       {Similar to Figure~\ref{distro_rojo_1}, but for the last position in 
       Figure~\ref{evol_rojo}. \label{distro_rojo_2}}

\section{Inversions}\label{sec_inv}

We have used the SIRJUMP code to invert the observed Stokes profiles. 
Based on SIRGAUSS \citep{Bellot-Rubio:2003fk}, this code can handle 
discontinuities or jumps in all the atmospheric parameters or some of 
them. The discontinuities are located at the same optical depth for the 
selected parameters. Initially, the amplitudes of the jumps are provided 
by the user, and then modified by the code until the differences between 
the observed and synthetic profiles are minimized. 
Examples of the application of this code to Hinode spectropolarimetric  
measurements can be found in \citet{Louis:2009lr}. 

In this paper, we show that one-component model atmospheres
with discontinuities of $|{\mathbf B}|$  and the
 line-of-sight velocity ($v_{LOS}$)
can explain the observed blue-only and red-only profiles. To that
end, we have inverted the profiles displayed in Figure~\ref{ejestks}
  using a fixed magnetic filling factor of 1, i.e., assuming that the whole
pixel is occupied by the same atmosphere showing discontinuous
stratifications.  
These profiles come from the two positions marked with yellow crosses in 
Figure~\ref{comparaI}. The colored dotted line  in Figure~\ref{ejestks} 
displays the best-fit profiles returned by SIRJUMP. Figures  \ref{inv_azul} 
and \ref{inv_rojo} show the temperature, $v_{LOS}$, the strength and 
the inclination of the magnetic field that the inversion code gave as solution. 
For simplicity, we only show here the observed and best fit Stokes I and V, 
although the inversion was done taking into account the four Stokes 
parameters as one can see in Figure~\ref{ejestks}. 
The vertical dashed lines delimit the region $-2 < \log(\tau) < 0$, 
where \ion{Fe}{1} 6301 \AA\ and 6302 \AA\  are more sensitive 
to perturbations of the atmospheric parameters \citep{Bas92, Bas94}.

For the blue-only profiles (Figure~\ref{inv_azul}) we retrieve a sharp 
discontinuity around $\log(\tau) = -0.7$. Below the discontinuity, the 
magnetic field has a strength of some 270~G and is rather inclined, 
making 120$^\circ$ to the vertical. In the same layers, we observe 
upflows of 2.5 km $s^{-1}$. Above the discontinuity, both the 
magnetic field strength and the LOS velocity are reduced significantly, 
down to 150 G and 0 km $s^{-1}$. With an inclination of 89$^\circ$, 
the magnetic field becomes more horizontal and even changes polarity. 
The configuration deduced from the inversion of the full Stokes vector 
showing blue-only profiles corresponds to a relatively weak magnetic 
field which is ascending in the deep layers toward a nearly field-free 
region at rest in higher layers. 

For the \rop\  shown in Figure~\ref{inv_rojo}, the variation of these
physical magnitudes occurs basically in the region $-1.2 < \log(\tau) < -0.8$. 
In this region, the velocity changes by 1.9 km s$^{-1}$ and the strength 
and inclination of the vector field by roughly 95~G and  
31$^{\circ}$, respectively. In other words, in the lower part of the atmosphere 
there exists a magnetized atmosphere with a
strong downflow, while in the upper part the atmosphere 
contains unmagnetized plasma flowing upwards.

\unafig{width=\textwidth}{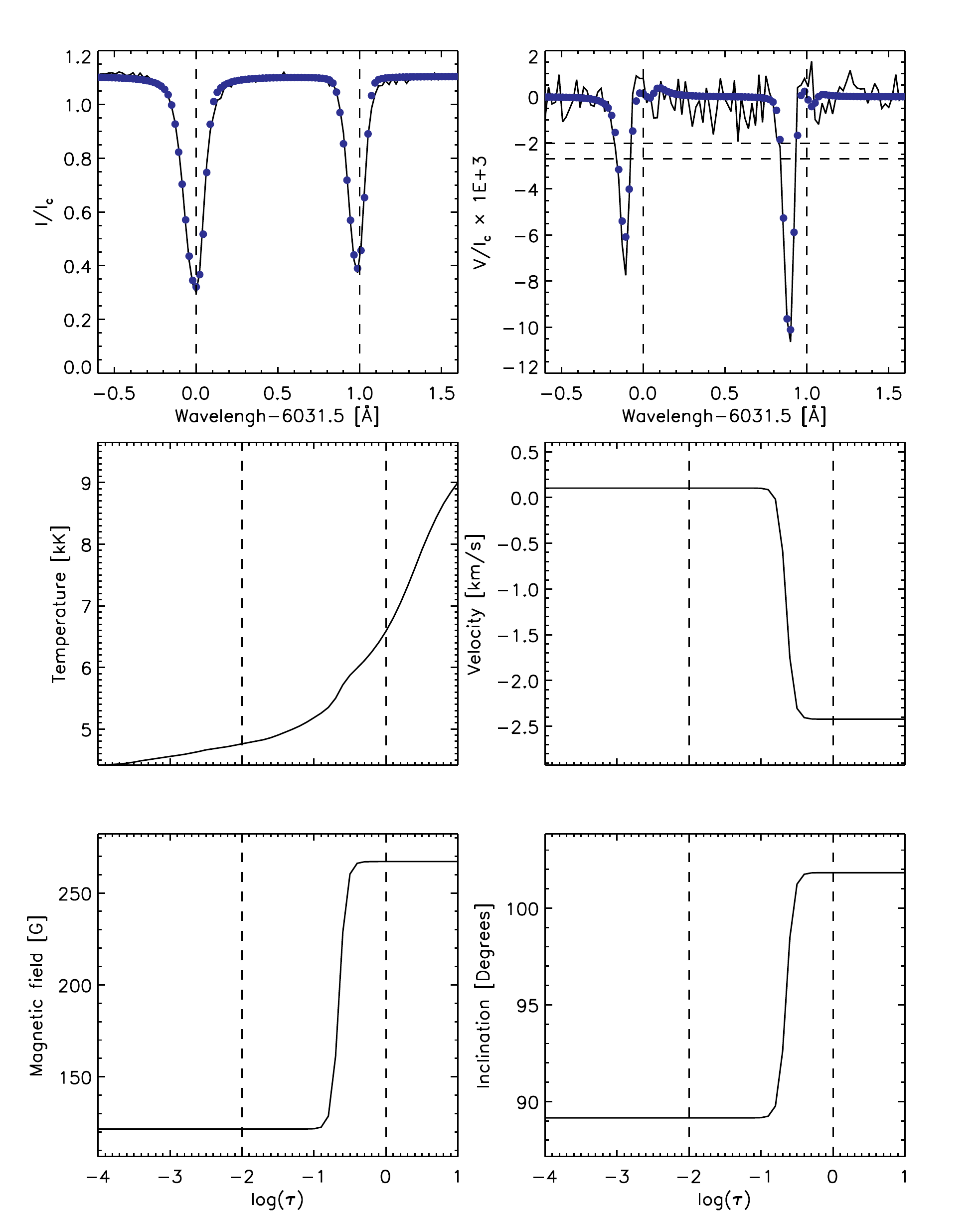}
	{SIRJUMP inversion of blue-only profiles. Top: observed and 
	best-fit Stokes I and V profiles (solid line and blue dotted line, respectively).  
	The observed profiles are those displayed in Figure~\ref{comparaI} 
	(the four Stokes profiles are shown in Figure~\ref{ejestks}). 
	Mid and bottom: temperature, velocity, magnetic field strength and 
	inclination retrieved from the inversion. Negative velocities indicate upflows. 
	As can be seen, all the atmospheric parameters show discontinuous stratifications
	within the line formation region.\label{inv_azul}} 

\unafig{width=\textwidth}{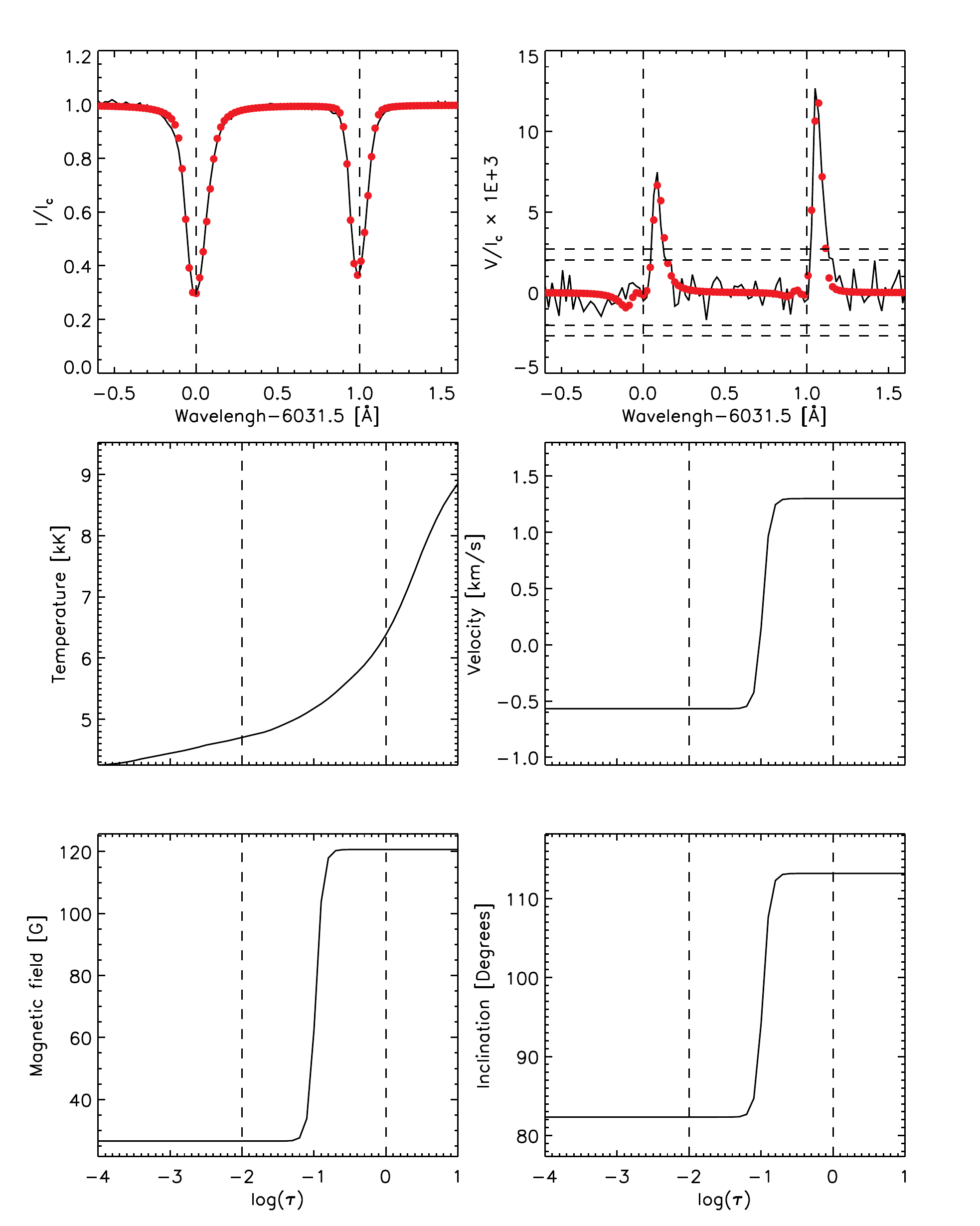}
	{Same as Figure~\ref{inv_azul}, for the \rop displayed in Figure~\ref{comparaI}. 
	\label{inv_rojo}}

The good fits provided by SIRJUMP demonstrate that the single
lobed Stokes V profiles can be explained with discontinuities in some 
atmospheric parameters. Both inversions show a stronger magnetic field 
in the lower layers than in the higher layers and rather 
inclined field lines in the lower part of the atmosphere.
Note that the strong field is emerging (submerging) 
for the \bop\ (\rop). This suggests that some of these profiles
might be related to the local evolution of the magnetic field within 
the granule. 

\section{Numerical simulation}\label{sec_numerical}

To gain a better understanding of the physical mechanisms behind the
\sls s and complement the inversions, we synthesized Stokes
profiles in \ion{Fe}{1} 6301 \AA\ and 6302 \AA\ from realistic 3D
radiative MHD simulations of flux emergence. The simulations were
run with the {\it Bifrost} code, which solves the MHD equations with
radiative transfer and conduction along the magnetic field
\citep{Gudiksen:2011qy}. The synthetic profiles were calculated
{\em a posteriori} using the SIR code and the modeled atmospheres
resulting from the simulation.

The computational domain stretches from the upper convection zone
(1.4~Mm below the photosphere) to the lower corona. We use a
non-uniform grid of $512\times 256\times 325$ points spanning
$16\times 8\times 16$~Mm$^3$, which implies a horizontal grid spacing
of $32$~km. The frame of reference for the model is chosen so that $x$
and $y$ are the horizontal directions.  The grid is non-uniform in the
$z$-direction to ensure that the vertical resolution is good
enough to resolve the photosphere and the transition region with a
grid spacing of $28$~km, while becoming larger at coronal heights
where gradients are smaller.

The initial model is seeded with magnetic field, which rapidly
receives sufficient stress from photospheric motions to maintain
coronal temperatures ($T>500\,000$~K) in the upper part of the
computational domain, as first accomplished by
\citet{Gudiksen+Nordlund2004}.  In the photosphere, the unsigned
initial magnetic field has a strength of $160$~G and is distributed in
two ``bands'' of vertical field centered around roughly $x=7$~Mm and
$x=13$~Mm. As a result, the magnetic field lines expand into the
corona forming loops oriented roughly along the $x$-axis.

To simulate the emergence of magnetic flux, we introduced a
non-twisted flux tube into and through the lower boundary of the
model. Specifically, horizontal fields of strength $10^3$~G were 
injected in a band of 1.5~Mm wide parallel to the $y$-axis and 
centered at $x = 8$~Mm through the bottom boundary, 1.4~Mm
below the photosphere \citep[for details, see the simulation labeled B1
in][]{Martinez-Sykora:2009rw}. Hence, the injected magnetic field is
nearly perpendicular to the orientation of the pre-existing ambient
field outlined by the coronal loops. This configuration is meant to
provide a rough representation of actual conditions in the quiet sun,
where flux is emerging continually into a preexisting magnetic field.

\begin{figure}[ht!]
    \includegraphics[bb = 37 0 757 368, clip=true, width=0.5\textwidth]{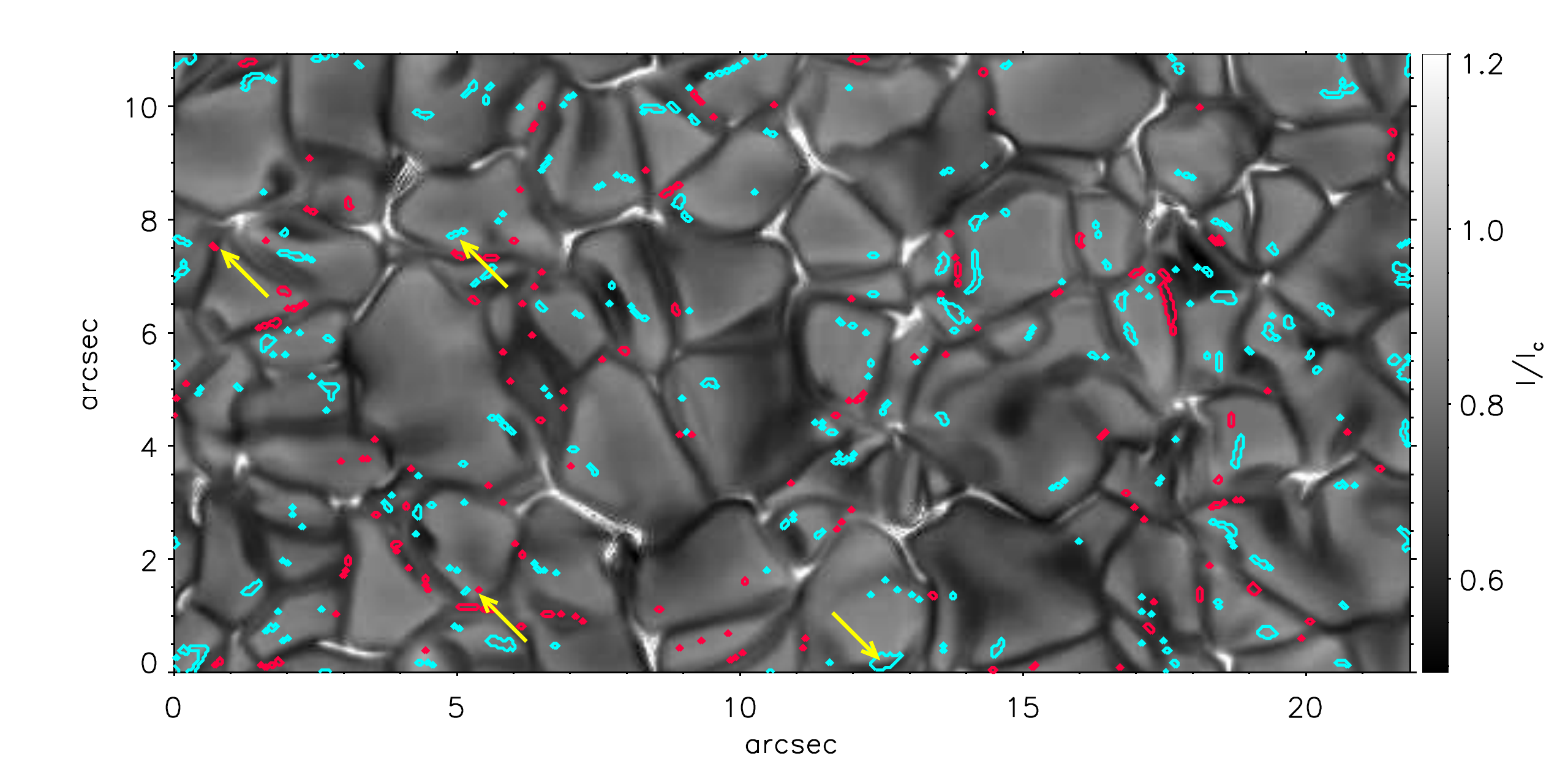}
  \caption{\label{fig:synlobs}Synthetic intensity map is shown
  at the photosphere in grey scale. Blue-only and \rop s locations are
  overplotted with blue and red patches respectively.}
\end{figure}

To compute synthetic Stokes profiles we considered each column of the
simulation as an independent one-dimensional atmosphere. The physical
parameters were interpolated to an evenly-spaced logarithmic optical
depth grid with stepsize $d\log \tau=0.03$. To identify \sls s we
applied the code described in section~\ref{sec_analisis} to the
synthetic profiles with $\sigma = 8\times10^{-4} I_c$.  We did not 
degrade the synthetic profiles to the spatial or temporal resolution of Hinode. 
However, the profiles were convolved with the spectral PSF and resampled 
to the Hinode wavelength scale.  We have also synthesized the Stokes 
profiles for different $\mu$ values: 1.0, 0.9, 0.8, 0.7, 0.6, 0.4 and 0.2. That
is, we have considered rays passing through the simulated models with
an inclination to the vertical of roughly $0^{\circ},\ 
26^{\circ},\ 37^{\circ},\ 46^{\circ},\ 53^{\circ}$ and 78$^{\circ}$.

The simulation shows many examples of blue-only and \rop s (see blue
and red patches in Figure~\ref{fig:synlobs}). In agreement with the
observations, the \rop s are mostly located in the intergranular lanes
and the \bop s are mostly centered in the outer part of the granules,
although the patches are slightly smaller than the observed ones.

The fraction of single-lobed profiles as a function of heliocentric
angle is presented in Figures~\ref{b_texp}, \ref{r_texp}
and~\ref{br_texp} with black diamonds.  As can be seen, the red-only
and \bop s are slightly less frequent than in the observations. This
could be due to several reasons: i) we did not degrade the profiles to
the Hinode spatial and temporal resolution; ii) the vertical spatial
resolution of the model may be not good enough to produce sharp jumps
in the photosphere; iii) the magnetic field configuration might be too
simplistic; iv) the amount of magnetic flux emergence in the box is
probably different from that occurring in the quiet sun. However, the
ratio of the blue-only to the \rop s seems 
to agree with the observations (Figure~\ref{br_texp}). Moreover, the
trend of the number of the occurrences of \bop s with $\mu$ is very
similar to the observations (Figure~\ref{b_texp}). For the \rop s, the
number of occurrences increases with $\mu$, and for $\mu<0.6$ it
remains almost constant (Figure~\ref{r_texp}).

In the simulations, we see that many different stratifications produce
single-lobed Stokes V spectra. These stratifications are often complex
because of the chaotic structure of the magnetic field in the
photosphere. However, most of them can be categorized in two broad
classes which show discontinuities or strong variations of the
physical parameters with height. The first group includes atmospheres
where the magnetic field is mostly concentrated in the deep layers.
The second group is that of canopy-like structures, i.e., a magnetized
region on top of a nearly-field free plasma.  Hereafter we will call
them ``deep-field'' and ``high-field'' configurations, respectively.

\begin{figure}[ht!]
    \includegraphics[width=0.5\textwidth]{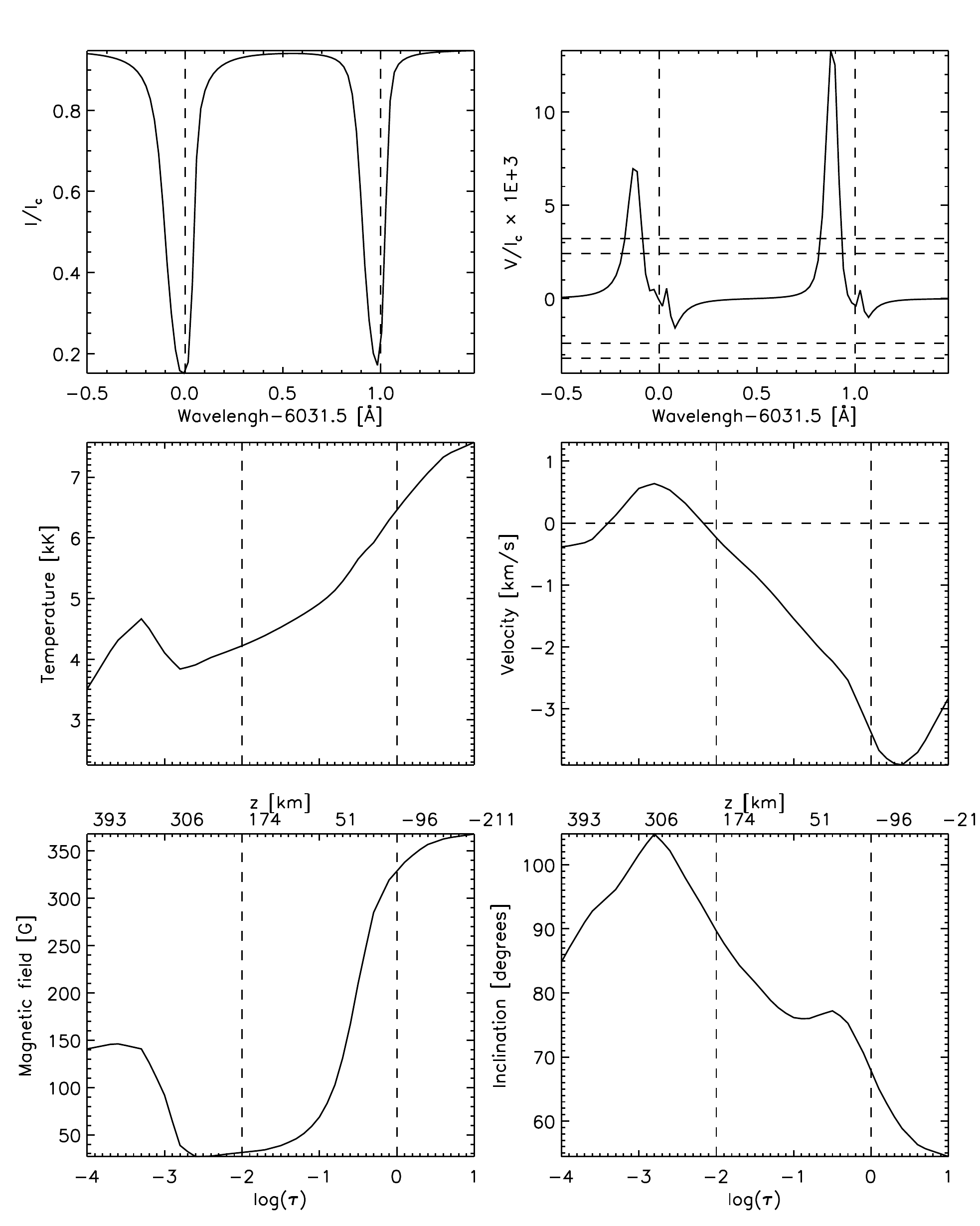}\\
        \includegraphics[width=0.5\textwidth]{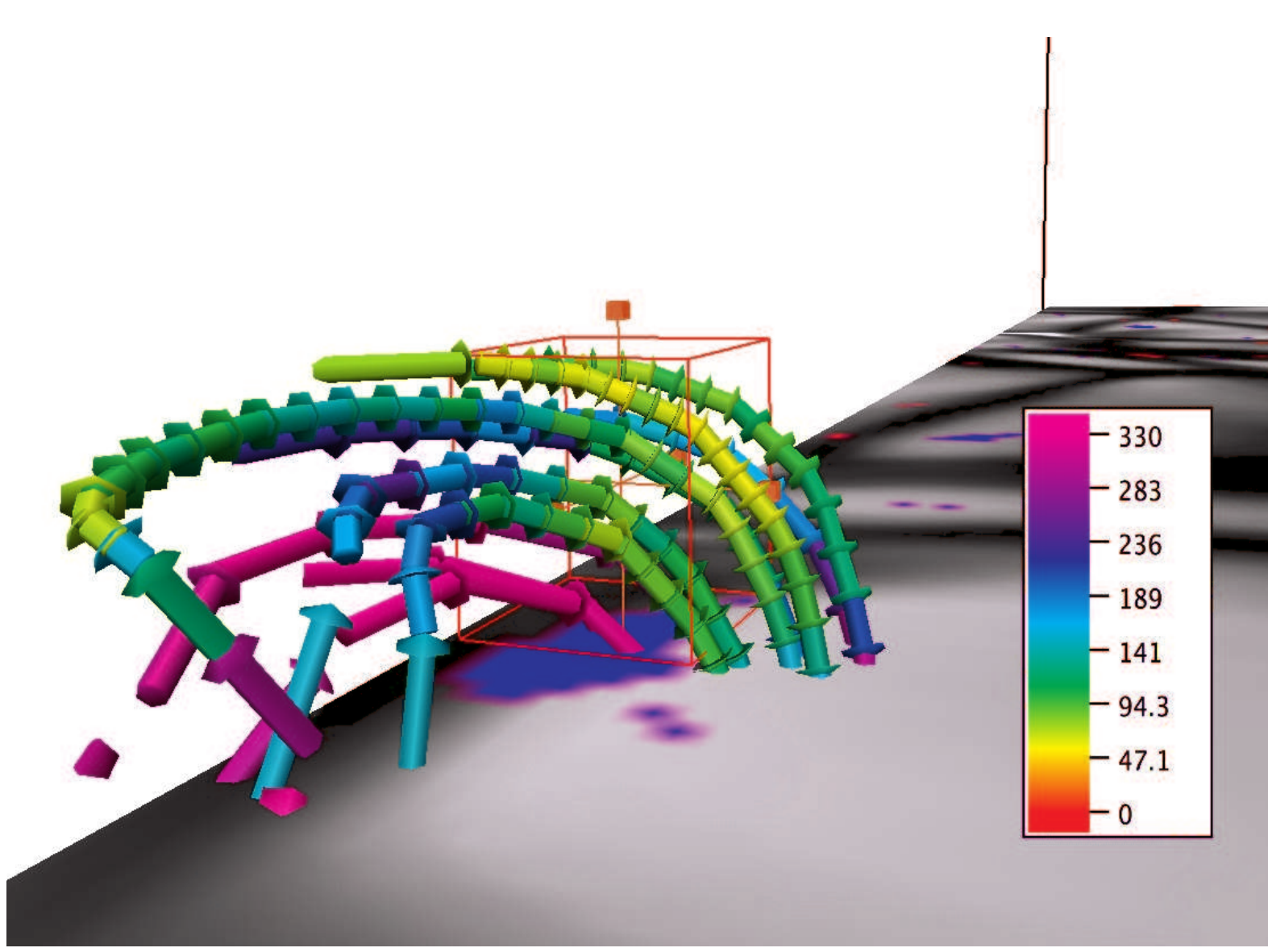}
 \caption{\label{fig:stratblue1} Blue-only profile observed in the simulation
at $[x,y]=[12.6,0.2]$~arcsec and the associated deep-field configuration
producing it. The layout of the three first rows is the same as in
Figure~\ref{inv_azul}. The synthetic spectra have been convolved
with the spectral PSF of Hinode and sampled with a wavelength step of
21.5~$m$\AA.  The horizontal dashed lines in the top-right panel are
drawn at the $\pm 3 \sigma$ and $\pm 4 \sigma$ levels. The bottom
panel displays a 3D view of the MHD model. The magnetic field is shown
with vectors where the color scheme represents the magnetic field
strength. The background image in grey is a continuum intensity
map. The red box indicates the optical depth range $ -2 < \log(\tau) <
0 $ for the pixel showing the blue-only Stokes V profile. The blue and
red patches mark the location of red-only and \bop s.}
\end{figure}

An example of the typical deep-field configuration for
\bop s is given in Figure~\ref{fig:stratblue1}. At the location
of the selected profile, the atmosphere exhibits a discontinuity with
a rather large jump in the magnetic field strength similar to that
obtained from the inversions. The field in deep layers is much
stronger than the ambient field above ($\sim$300~G vs $\sim$30~G). The
velocity shows a strong gradient instead of a jump, with large upflows
in the deep layers and nearly zero velocities in the high
layers. The magnetic field inclination varies only
slightly. This configuration, which is very similar to that inferred
from the inversion, is produced by a small-size magnetic $\Omega$ loop
emerging at the edge of a granule. The magnetic topology around the
location of the blue-only profile is shown in the bottom panel of
Figure~\ref{fig:stratblue1}. As can be seen, the magnetic field in
deep layers is stronger (purple in the figure) than the ambient field
above (yellowish) and exhibits upflows.  The $\Omega$-loop is strongly
bent at the photosphere and only one of its footpoints is located
inside the granule. As is usually the case, this blue-only profile is
located at the edge of the granule but not yet in the intergranular
lane. It lasts for about 11 minutes. 

\begin{figure}[ht!]
    \includegraphics[width=0.5\textwidth]{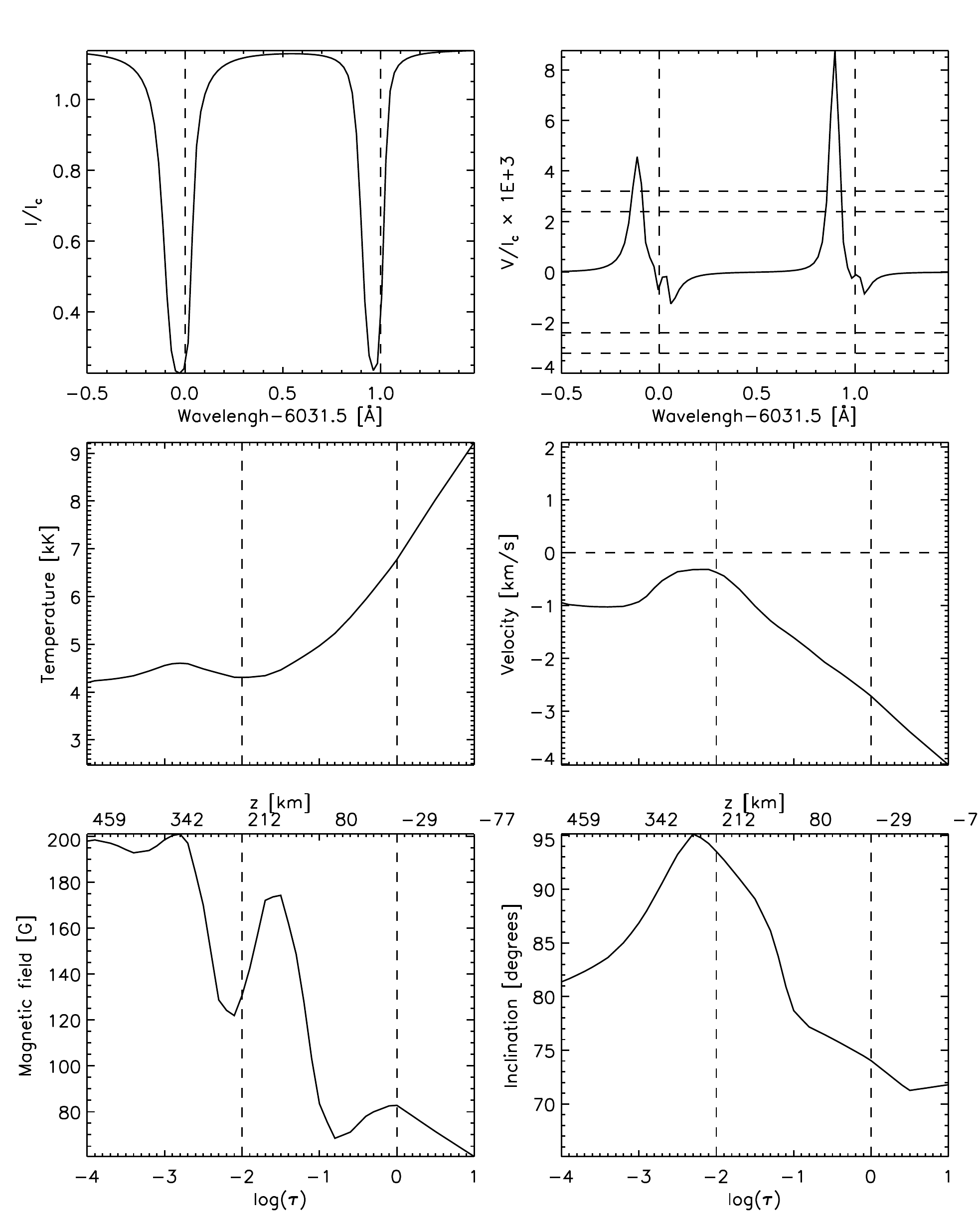}\\
        \includegraphics[width=0.5\textwidth]{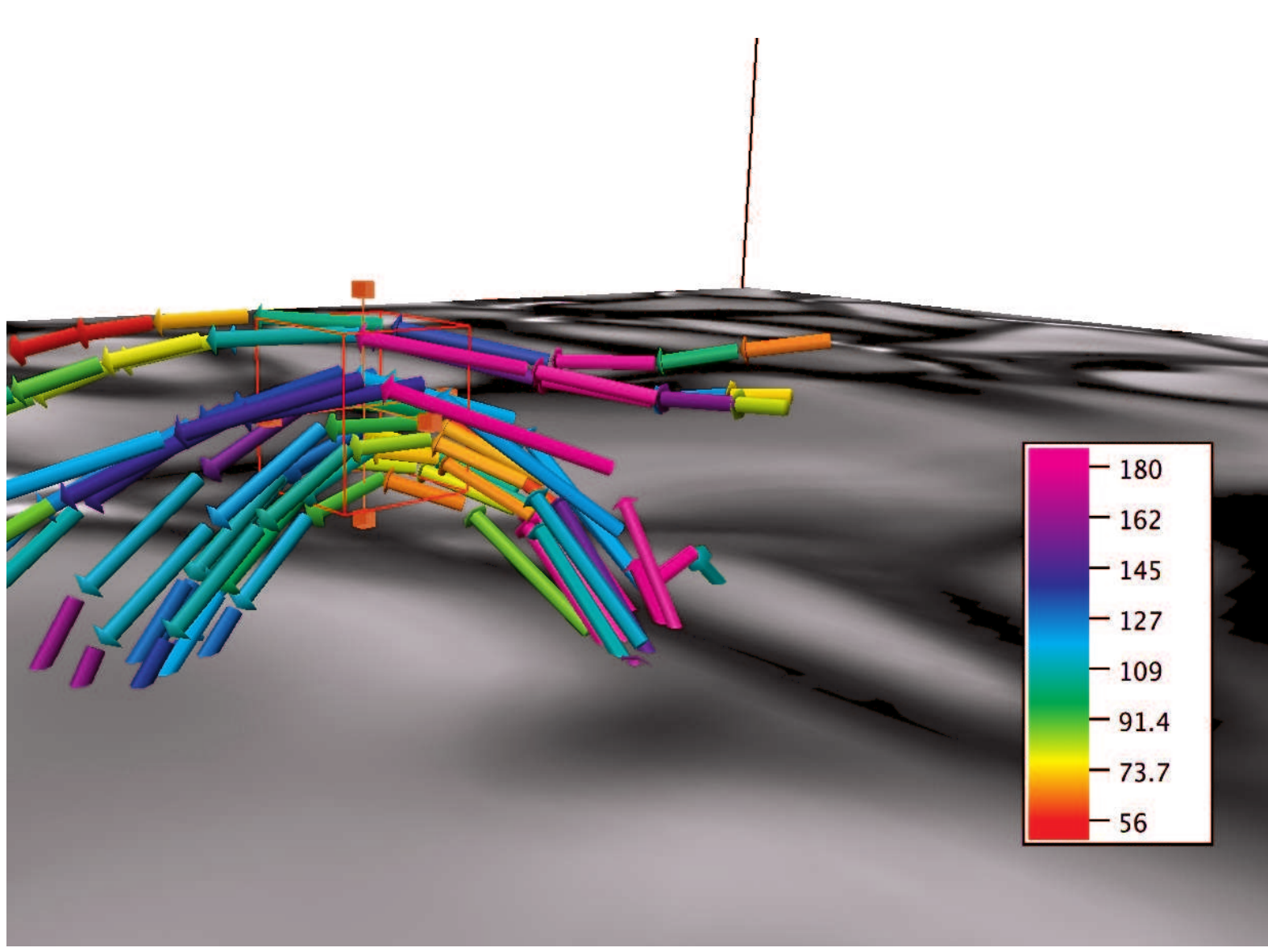}
\caption{\label{fig:stratblue2} Example of a \bop\ produced by a high-field
  (canopy-like) configuration at $[x,y]=[4.9,7.7]$~arcsec.
  The different panels are arranged as in Figure~\ref{fig:stratblue1}.}
\end{figure}

Blue-only profiles can also be produced by high-field (canopy-like)
configurations with upflows in the deep layers. In the example of
Figure~\ref{fig:stratblue2}, the lower part of the photosphere has
nearly zero magnetic fields ($\sim$80~G) and strong upward velocities
($\sim$-3~km~s$^{-1}$). Higher layers show stronger fields
($\sim$160~G) and nearly zero velocities ($\sim$-0.5~km~s$^{-1}$).
Only the magnetic field strength and inclination show abrupt jumps.
The velocity also varies quite strongly with depth, but in a more
linear fashion. Note that the gradient of magnetic field strength and 
cosine of the field inclination have opposite signs. In this case, it is the
variation of the inclination that determines the sign of the area 
asymmetry (i.e., the suppression of the Stokes V red lobe), because 
the magnetic field is nearly horizontal. In fact, most of the high-field 
configurations for the blue-only profiles show this type of stratification. 
Interestingly, the field strength attains a minimum in
the mid-photosphere, which suggests an interaction between two flux
systems--- possibly a reconnection process. As can be seen in the 3D
view at the bottom of Figure~\ref{fig:stratblue2}, the group of field
lines in the lower part of the volume has the shape of a small
$\Omega$-loop. The stronger magnetic field above is connected to other
magnetic structures further away. Looking at the time evolution
we observe that the upper field lines emerge and expand into the
chromosphere. The duration of this profile is also 11 minutes.

In the simulations, $\sim$85\% of the observed \bop s are due to
deep-field configurations associated with magnetic flux emergence at
the border of granules. About 65\% of the blue-only profiles show a
discontinuous magnetic field strength stratification like in
Figure~\ref{fig:stratblue1}, while $\sim$20\% exhibit a strong field
strength gradient instead of a pure discontinuity. The jump of the
different parameters is mostly located deep in the photosphere, inside
the optical depth range $ -1 < \log(\tau) < 0 $. A velocity gradient
of the type shown in Figure~\ref{fig:stratblue1} is present in most of
the \bop s ($\sim$95\%).

Another $\sim$10\% of the blue-only profiles are associated with
high-field, canopy-like configurations similar to that shown in
Figure~\ref{fig:stratblue2} (note the complexity of the magnetic
field). The vast majority (99\%) of the high-field configuration 
exhibits a discontinuity in the magnetic field strength. These 
discontinuities are usually located above $\log(\tau)=-1$.  

Finally, the remaining 5\% of \bop s are
the result of complex atmospheric stratifications which do not fit in
these two classes. Irrespective of their origin, all the blue-only profiles show
large upflows in deep photospheric layers and nearly zero velocities
higher up.

\begin{figure}[ht!]
    \includegraphics[width=0.5\textwidth]{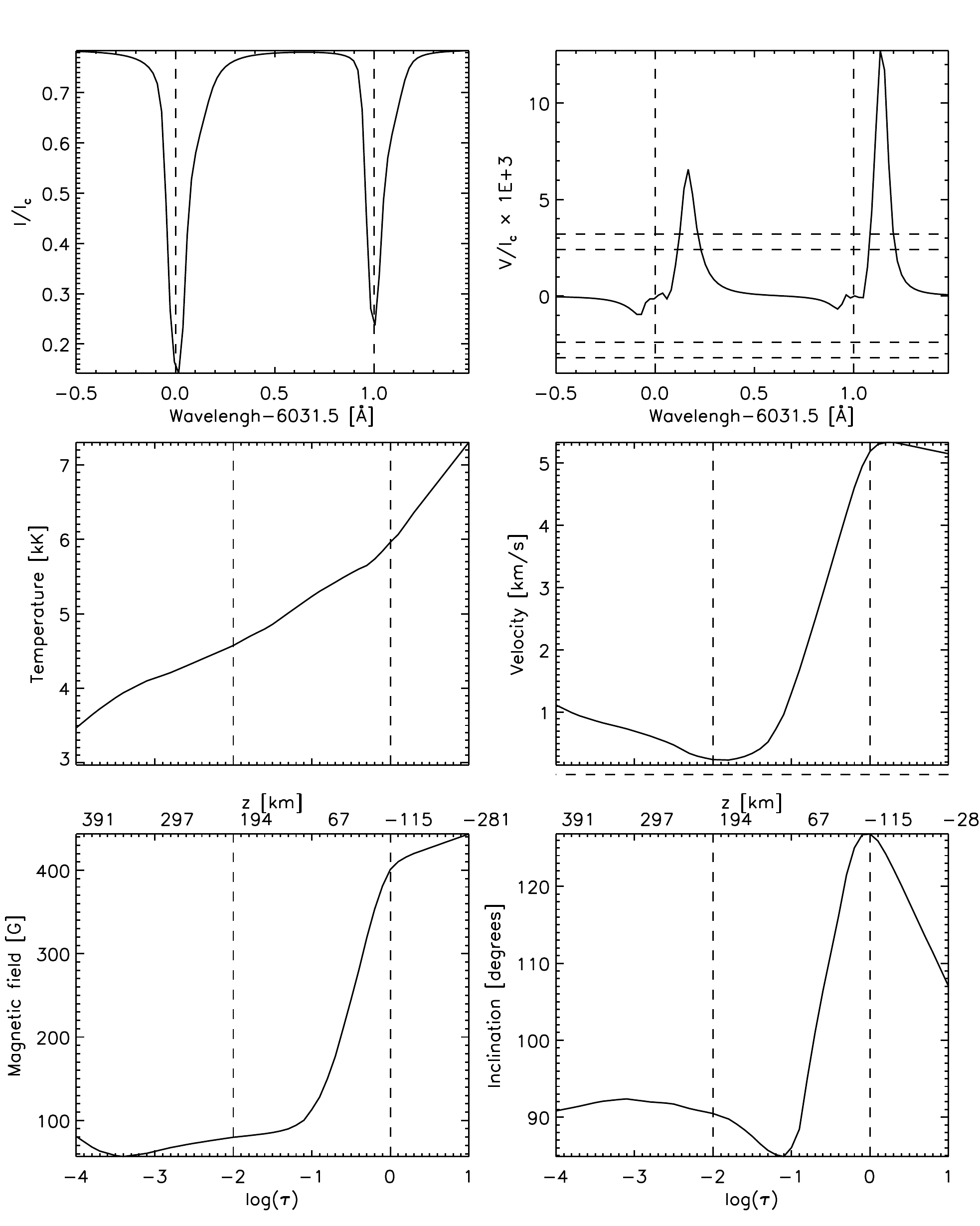}\\
        \includegraphics[width=0.5\textwidth]{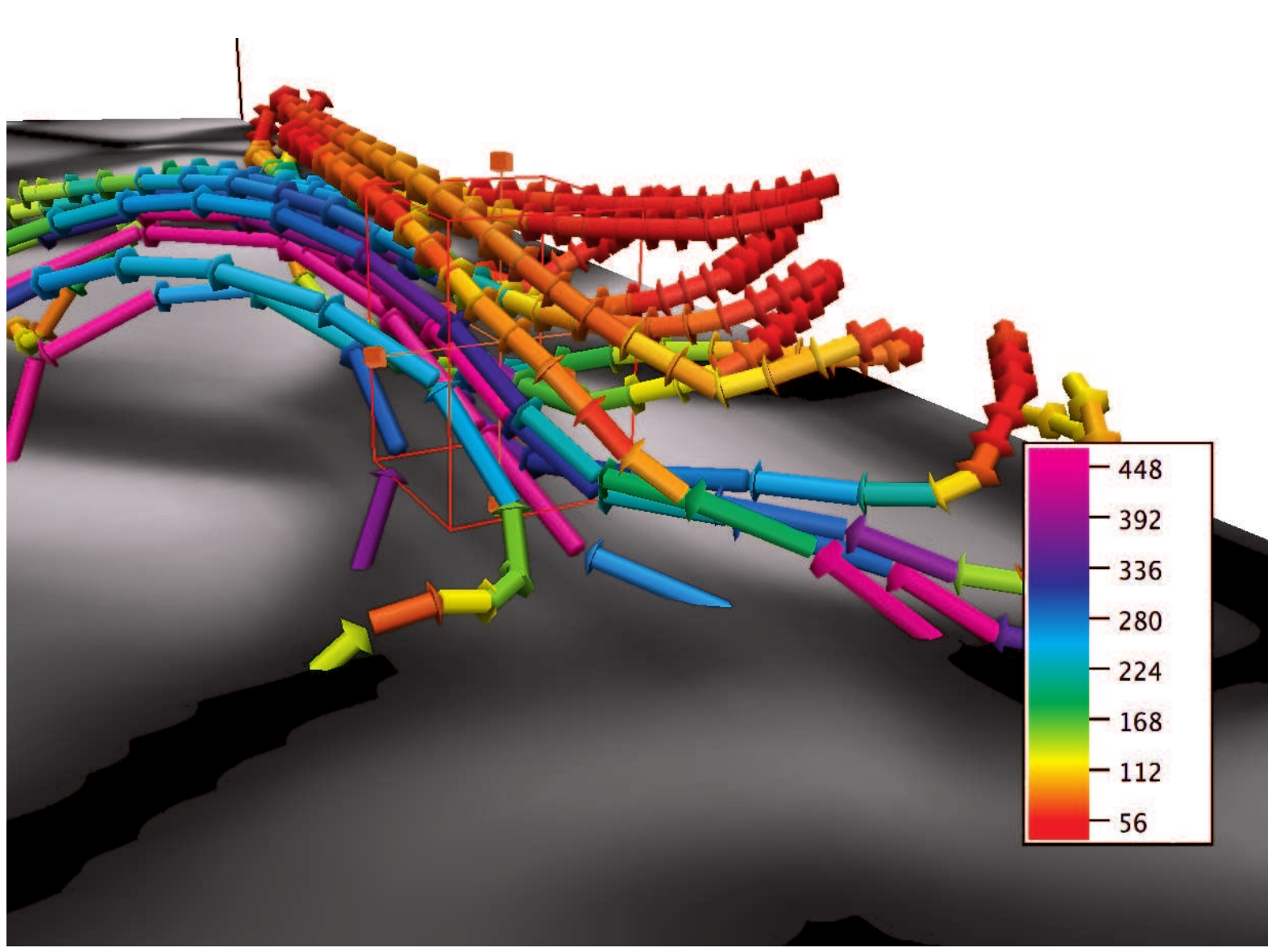}
  \caption{\label{fig:stratred} Example of a \rop\ arising from a deep-field
  configuration at the location $[x,y]=[5.5,1.1]$~arcsec. The arrangement
  of the panels is as in Figure~\ref{fig:stratblue1}.}
\end{figure}

We find that also the red-only profiles are produced by deep-field 
and high field configurations, this time associated with downflows. 
The most common source of \rop s is the deep-field configuration, 
a typical example is shown in Figure~\ref{fig:stratred}. This 
atmosphere exhibits a discontinuity in magnetic field, which is much 
stronger in deep layers ($\sim$400~G vs 100~G). The velocity
shows a strong gradient with large downflows in the deep layers
 ($\sim$5~km~s$^{-1}$ vs 0.5~km~s$^{-1}$). The field inclination
has a small jump from slightly inclined field in the lower layers to
horizontal field in the higher layers. This configuration is in agreement 
with the stratification obtained from the inversion of the \rop s of 
Figure~\ref{ejestks}. The rather strong and deep magnetic field is 
confined along the intergranular lane almost parallel to the
photosphere as shown in Figure~\ref{fig:stratred}. These lines
end into the photosphere as an $\Omega$-loop. Above them, another
set of field lines, also following the path of the intergranular lane,
have weaker magnetic strength and are connected to a different
place in the photosphere. The deep magnetic field is submerging into 
the intergranular lane. The duration of this \rop\ is roughly 12 minutes.

\begin{figure}[ht!]
    \includegraphics[width=0.5\textwidth]{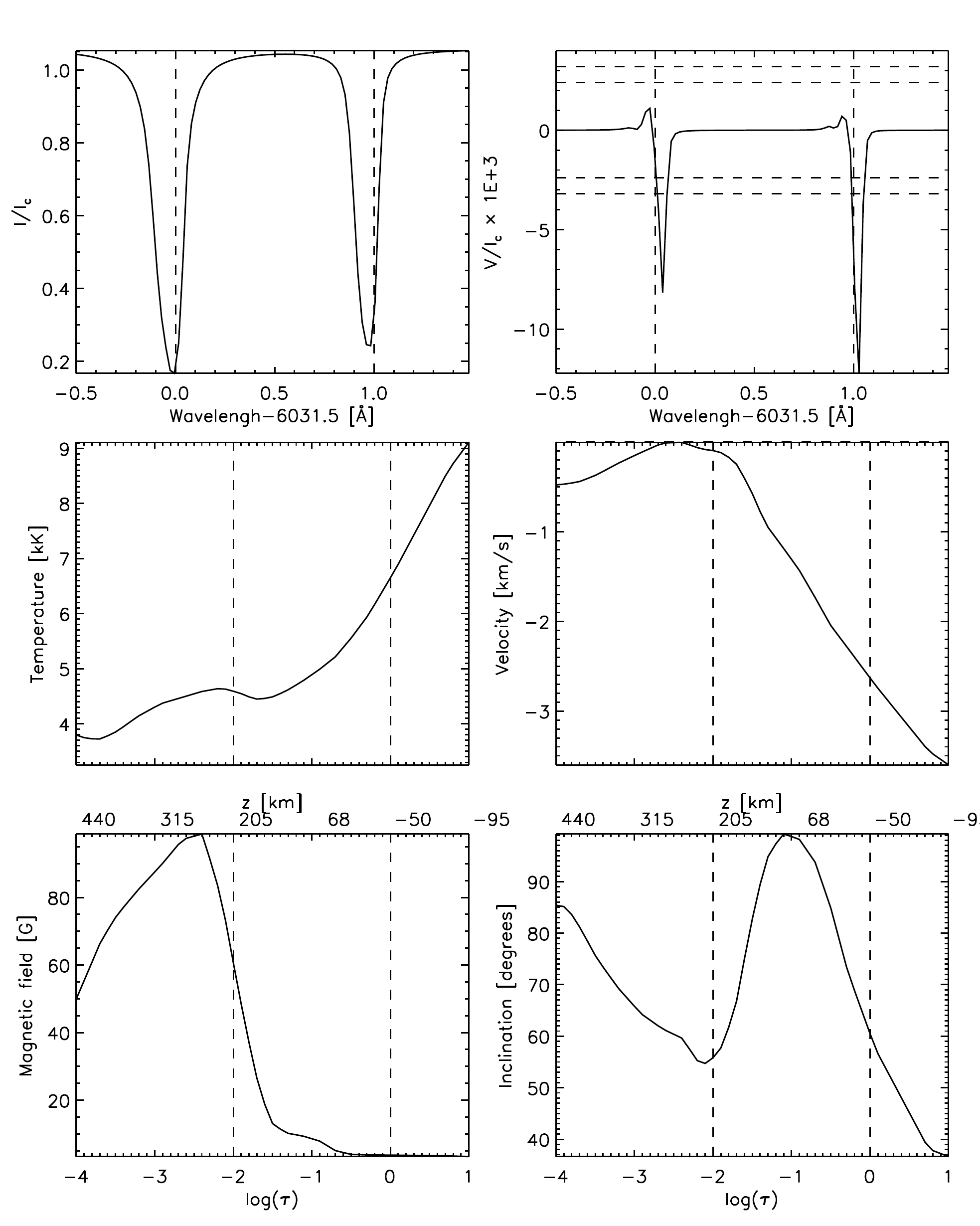}\\
            \includegraphics[width=0.5\textwidth]{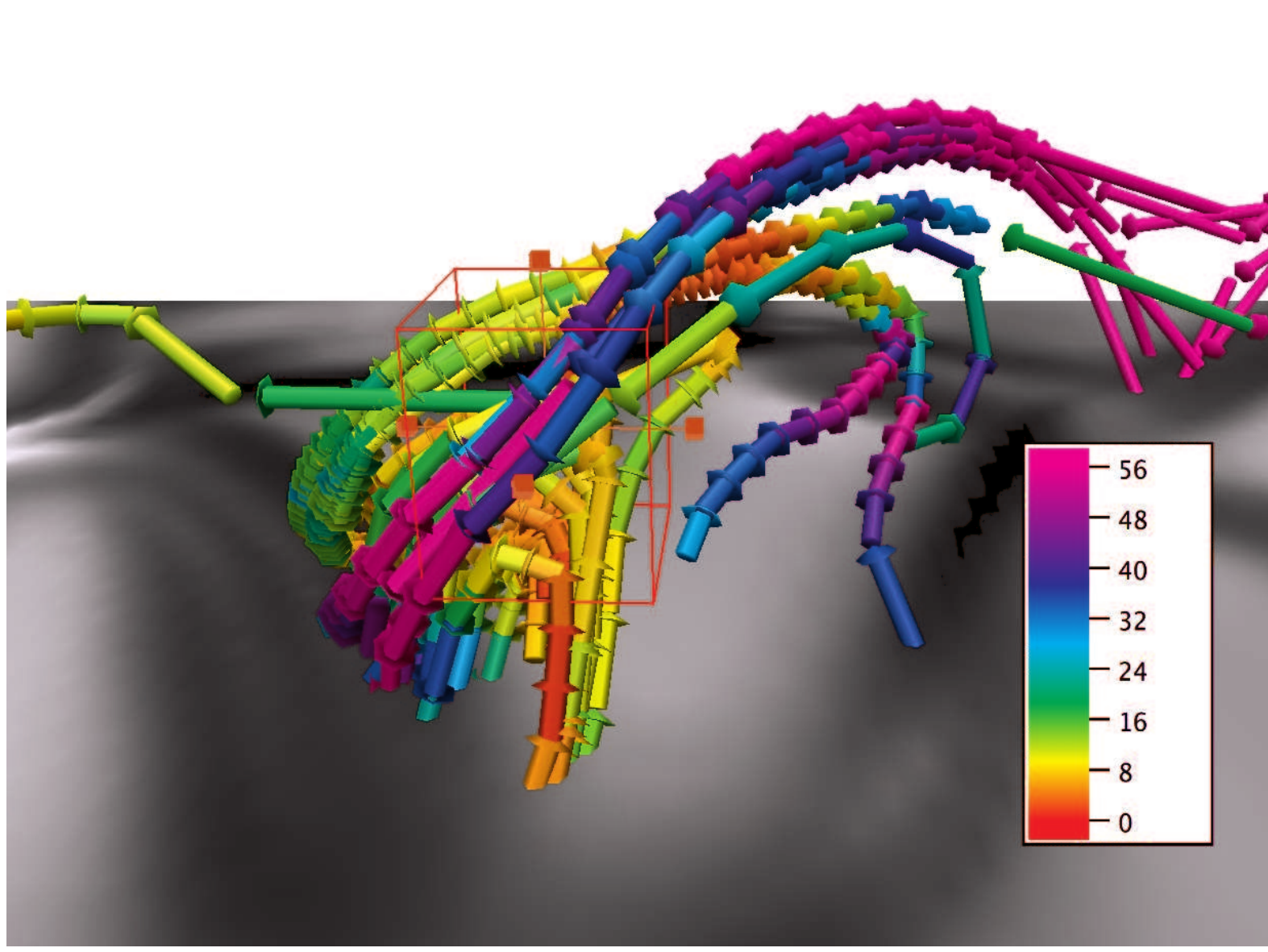}
  \caption{\label{fig:stratred2} Red-only profile  due to a high-field configuration at
  $[x,y]=[1.9,6.7]$~arcsec. The arrangement of the panels is as in
  Fig~\ref{fig:stratblue1}.}
  \end{figure}

An example of a high-field configuration 
leading to red-only profiles is displayed in Figure~\ref{fig:stratred2}. 
For these profiles to appear it is necessary that the deep
photosphere has weaker magnetic field and stronger
upward velocities than the high photosphere ($\sim 5$~G vs 80~G
and $\sim$~-3.5 vs -0.4~km~s$^{-1}$, respectively).  
The magnetic field strength and inclination show 
a rather strong jump, while the LOS velocity exhibit a 
steep gradient. This configuration is a clear example 
of a canopy atmosphere. Note that this particular 
\rop\ is located inside the granule as shown in the 3D image in the bottom
panel of Figure~\ref{fig:stratred2}. The group of field lines
located in the lower photosphere has the shape of a tiny
$\Omega$-loop. These tiny loops show a strong turn over
at the height where the strong magnetic fields are located higher 
up. The lifetime of this profile is roughly four minutes, because the 
``canopy'' appears only during the splitting in two parts.

In summary, in the simulation most of the \rop s (82\%) are associated
with magnetic fields that are submerging (as the example shown
in Figure~\ref{fig:stratred}). About $\sim$70\% of them
have a discontinuity in the magnetic field strength and $\sim$15\% 
show a smooth decrease of the magnetic field strength with height.
Sometimes, the field strength reaches a minimum in between the 
ambient and the submerging field, which suggests an interaction between 
the two flux systems, possibly a reconnection process (20\% of the cases).
About $\sim$15\% of the \rop s are produced by a high-field configuration such 
as that of the Figure~\ref{fig:stratred2}. Finally, the remaining
2\% of \rop s are the result of complex atmospheric stratifications
which do not fit in these two classes.

Thus, the stratifications that lead to \sls s in the simulations support
the main conclusion derived from the inversion, namely that jumps in the 
physical parameters along the line of sight can explain the existence of 
single lobed profiles. This is not the only mechanism capable of producing 
such anomalous signals, since other types of stratifications also generate
circular polarization profiles with only one lobe, but it is the most common 
as observed in the simulations. 

From the many stratifications producing the single-lobed profiles in the simulations,
when the stratification is simple enough, we note that if the variation of
the velocity with height is opposite (equal) to the variation of the magnetic field 
strength with height, the single-lobed Stokes V is blue-only (red-only) in 
accordance with the rules deduced by \citet{Sol88} and \citet{Ste99}. 
This is not the case for the example of Figure~\ref{fig:stratblue2} because 
in that case the important parameter is the gradient of the cosine of the 
magnetic field inclination. 

\section{Summary}\label{sec_sum}

Using Hinode spectropolarimetric measurements, we have 
performed an in-depth characterization of the circular polarization 
profiles with only one lobe observed in the quiet sun network 
and internetwork. We have successfully inverted them in terms 
of model atmospheres featuring discontinuities in the parameters 
along the line of sight. The scenario indicated by the inversions 
is supported by the results of 3D MHD simulation which also 
show an abundance of blue-only and red-only Stokes V profiles. 

The main properties of these profiles can be summarized as follows:
\begin{itemize} 
\item[-] The \bop s are mainly located in the outer part of the granules, 
therefore they are more often related with upflows. They form 
patches as large as $1\arcsec \times 1\arcsec$ and their mean 
lifetime is slightly shorter than the granule lifetime.  
\item[-] The \rop s are mainly located in the intergranular lanes, 
therefore they are more often related with downflows. 
They tend to form very small structures, sometimes as small 
as a single pixel, i.e., $0.15\arcsec \times 0.16\arcsec$. 
\item[-] The ratio of blue-only to red-only profiles  
drastically decreases from the center to the limb of the solar disk. 
This variation can be traced back to a strong reduction of the 
number of blue-only profiles. Such a reduction may be caused 
by the upward shift of the $\tau = 1$ level toward the limb, 
which makes it difficult or even impossible to reach the deep 
layers where these ``granular" profiles are formed. 
\item[-] The area occupied by the \sls s on the solar surface is less 
than 2\% ($\sim$1\% of the area is occupied by blue-only profiles and 
$\sim$0.4\% by red-only profiles). 
\item[-] However, the fraction of blue-only and red-only profiles, 
compared to the profiles that show clear signal in Stokes V, 
are 4\% and 1\%, respectively.
\item[-] The \sls s are located in the surroundings of network 
patches, never inside them. They are often located in places where 
there are simultaneously both circular and linear polarization 
signals, which suggests that they are associated with inclined 
fields, or that they occur because of a sudden change in the 
direction of the magnetic field. 
\item[-] Both the blue-only and red-only profiles show weak or 
modest circular polarization signals and weak linear polarization 
signals. 
\item[-] The evolution of the \sls s is related to the evolution of 
the granules and intergranular lanes that host them. 
\item[-] Inversion of the four Stokes profiles based on models featuring 
discontinuities along the LOS can reproduce the observed \sls s. 
\item[-] Synthetic single-lobed profiles are observed in realistic 3D 
numerical simulations. Several types of stratifications can explain 
these profiles, but most of them show discontinuities in at least one of 
the parameters (usually the magnetic field strength or the inclination)
as was inferred from the inversion. A large fraction of these stratifications 
show a strong velocity gradient instead of a jump. 
\item[-] The number of single-lobed profiles and the statistics from the 
simulations are quiet similar to the observations, but there exist small differences
that might be caused by missing physics (see Section~\ref{sec_con}).
\end{itemize}

\section{Discussion and Conclusions}\label{sec_con}

The statistical and physical properties of the \sls s tell us that they 
cannot be explained as the superposition of opposite-polarity 
Stokes V profiles or as unusual events on the solar surface. On the contrary, 
they show clear features associated with physical processes that could 
involve a sharp or drastic change of the orientation and 
strength of the magnetic field occurring mainly between the outer 
part of granules and the adjacent intergranular lanes.

From the few temporal evolutions that we studied, the 
\bop s suggest a magnetic flux emergence  
processes, with lifetime comparable to the granulation lifetime.
Therefore, these patches first appear with weak polarization signal, then 
increase in size and strength. Finally, the blue-only 
patches disappear, but small Stokes V signal remain in 
the region where the blue-only patches were sited for roughly one minute.
We do not detect linear polarization with this profiles 
At the end, the signal in Stokes V disappears with the disappearance 
of the granule, perhaps as result of a submergence process. 

We used the SIRJUMP inversion 
code on this study after selecting suitable cases. This
code can handle strong variations of the vector magnetic 
field and velocity along the line of sight, i.e., it can model a 
magnetopause or slab structure in the pixel. In fact, both the inversion 
and the numerical simulations indicate that single-lobed profiles can be
produced by discontinuities along the line of sight. In the simulations a wide
range of atmosphere give rise to Stokes V profiles with only one lobe. 
However, most of them are related to emerging magnetic field (blue-only) 
and disappearing or submerging magnetic field (red-only).
Therefore, the statistical comparison between observations and simulations
provides a tool to determine both the magnetic field
topology in the quiet sun and its emerging and disappearing process in the local 
granulation. The lifetime of the \sls\ must tell us the duration of the 
emergence and the submergence, i.e., the duration of the fundamental 
input and output processes of magnetic flux. 

The absence of \bop s near the limb might be due to these small 
structures being located in the lower-middle atmosphere 
while the \rop s might be located more or less equally at different heights
or with some preference higher layers (this is supported 
by what we observed in the inversions and in the forward model).

As mentioned above, the simulation has a small discrepancy with the observations
in the statistics. This discrepancy could be due to the following reasons: 
\begin{itemize} 
\item[-] The vertical spatial resolution of the simulations is not good enough.
\item[-] The spatial resolution of Hinode was not taken into account.
\item[-] The amount of mean magnetic field, or the simplified initial magnetic 
configuration, may be off.
\item[-] It is probably necessary to input periodically a specific amount 
of new emerging magnetic flux in the simulation. The input of new flux might 
require a different type of magnetic field configuration in the bottom boundary. 
Moreover, the duration of each patch with new flux should last as long as the 
mean lifetime of the \bop s (as shown in Section~\ref{sec_temp}).
\end{itemize} 
Once the simulations have good grid resolution and the 
synthetic profiles have been convolved with the spatial and spectral PSF, 
then any remaining discrepancies between observations and
simulations must tell us with which frequency, where, and how long new and 
small structures of magnetic field have to be injected. In order to approach a realistic 
simulation of the convection zone and photosphere of the quiet sun, the simulations
must to reproduce the statistic study presented here (Sections~\ref{sec_spatial}-\ref{sec_solardisk}). 

We have focussed our attention on the quiet sun, but plage, active regions, and 
ephemeral regions might behave differently. Thus, it is important to carry out similar 
studies in those structures for a better understanding of their magnetic topology and 
evolution.

\section{Acknowledgments}
This work was partially carried out while ASD was a Visiting Scientist at the 
Instituto de Astrof\'isica de Andaluc\'ia. Financial support by the 
Spanish MICCIN through projects  AYA2009-14105-C06-06 and 
PCI2006-A7-0624, and by Junta de Andaluc\'ia through project 
P07-TEP-2687 (including a percentage from European FEDER funds) 
is gratefully acknowledged. Part of the data used here were acquired 
in the framework of HOP 14 (Hinode-Canary Islands joint campaign). 
Hinode is a Japanese mission developed by ISAS/JAXA, with NAOJ as
domestic partner and NASA and STFC (UK) as international partners. It
is operated in cooperation with ESA and NSC (Norway).
The Hinode project at Stanford and Lockheed is supported by NASA 
contract NNM07AA01C (MSFC).
ASD thanks the International Space Science Institute (hereafter ISSI) and I. Kitiashvili for the 
invitation to participate in the meeting `Filamentary Structure and Dynamics 
of Solar Magnetic Fields,' 15-19 November 2010, ISSI, Bern, where many aspects 
of this paper were discussed with other colleagues. 
This work has also benefited from discussions in the Flux Emergence meetings 
held at ISSI, Bern in February and December 2011. 
We thank T. Tarbell for his 
corrections and suggestions. 
The 3D simulations have been run with the Njord and Stallo cluster from the Notur
project and the Pleiades cluster through computing grants SMD-07-0434, SMD-08-0743,
SMD-09-1128, SMD-09-1336, SMD-10-1622, SMD-10-1869,  SMD-11-2312,
and  SMD-11-2752 from the High
End Computing (HEC) division of NASA. 
We thankfully acknowledge the computer and supercomputer 
resources of the Research Council of Norway through grant 170935/V30 and through 
grants of computing time from the Programme for Supercomputing. 
To analyze the data we have used IDL and Vapor (http://www.vapor.ucar.edu).


\clearpage
\appendix
The first seven columns in Table \ref{latabla} show parameters 
for all the observation runs. The spectral sampling was 
21.5$\pm$0.1 $m$\AA\ for all of them. 

\begin{figure}[h!]
	\begin{center} 
	\includegraphics[width=14cm]{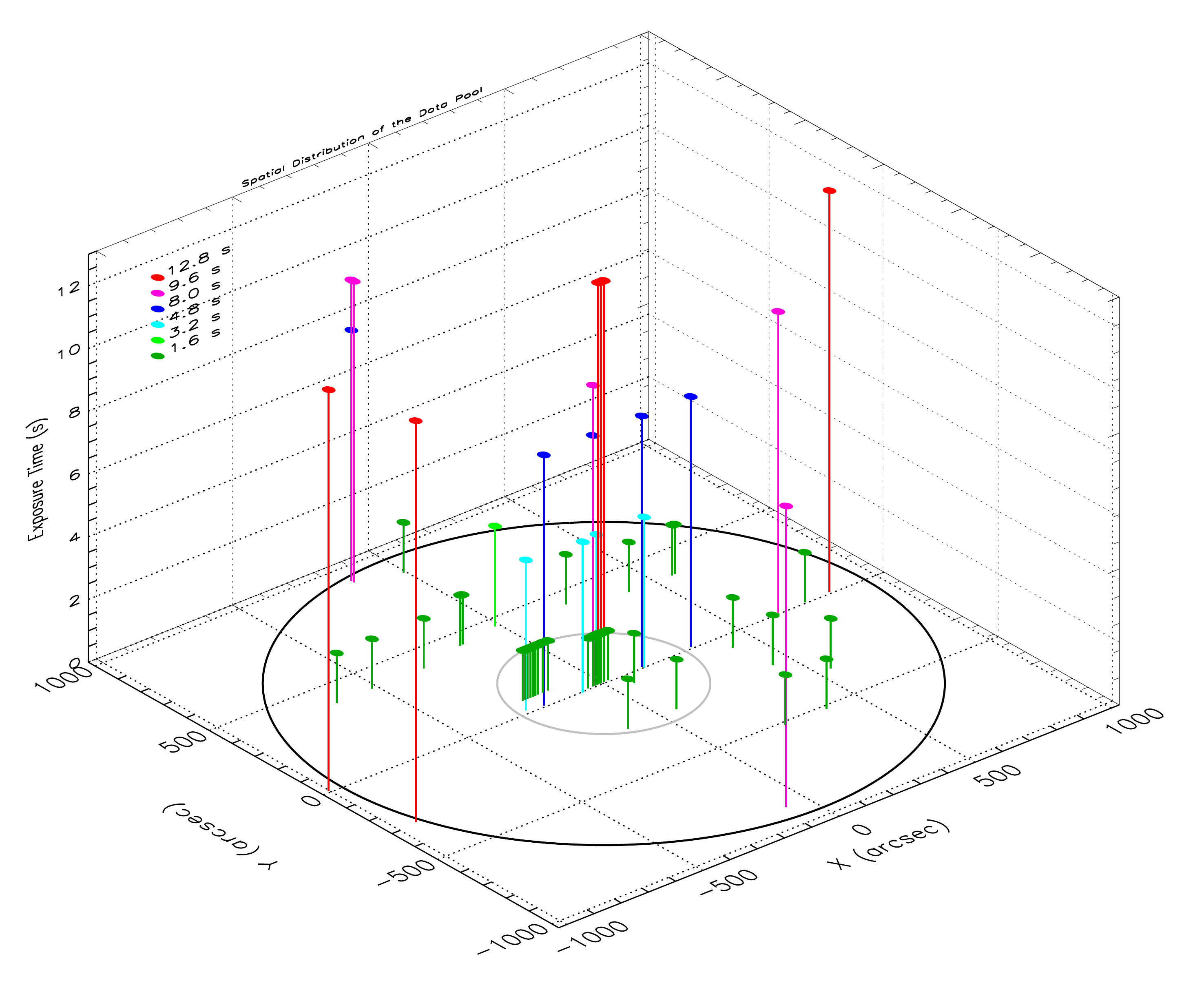}
	\caption{Distribution of the data pool on the solar disk. In the 
	inner grey circle are 61\% of the studied observation runs, that 
	we have called {\it population I}. They were observed with
	a wide range of exposure time. Out of this inner grey circle 
	stays the {\it population II}. They were mainly observed with 
	long time exposure time.\label{pos_texp}}
	\end{center}
\end{figure}

Figure \ref{pos_texp} shows the spatial distribution of the 
studied observational runs: the data pool. In the plane XY is 
drawn the solar disk (both axis are given in arcsec). The 
drawing pins locate the position of the selected 
72 Hinode SOT/SP observation run. 
The length of drawing pins represents the exposure time of 
the observation. In addition, we have 
used different colors for each exposure time.

We have selected 72 observation runs with different 
exposure time. 61\% of the observations are located at $1.00 < \mu < 0.95$, 
which are inside of the inner grey circle (see Figure~\ref{pos_texp}).
The observation runs inside the inner circle have a wide 
range of exposure times (1.6, 4.8, 8.0 9.6 and 12.8 s). 
For understanding the possible influence of the projection 
effect in the behavior of the single-lobed Stokes V profiles 
we have selected observations out of this centered region 
of the solar disk and in the limb (even two observation 
scans started very off the limb but ended inside the solar disk).
Therefore, we can talk about two populations or sets in our 
data pool: one is inside the inner circle (with a low 
influence of projection effect), hereafter called as {\it population I}, 
and the other is outside of this inner circle (with not negligible 
projection effect), which is referred as {\it population II}.  

Most of the scans were taken in quiet sun, both large area 
scans and temporal scans. We have also selected a few 
temporal scans taken over a part of sunspot or pore, but always including 
the photospheric network in their field of view. In addition, 
several scans close to the limb include points off the solar 
disk. These three kind of maps have been indicated in the last 
column of Table \ref{latabla} labeled {\it SS}, {\it P} and {\it OL} that 
stand for sunspot, pore and off the limb observations respectively. 
\nopagebreak
\nopagebreak
\setlength{\LTcapwidth}{7in}
\begin{center}
\begin{longtable}{ccccccccccc}
\caption[]{\textrm{Data Pool. From left to right: date and time, 
	spatial scales, position on the solar disk, exposure time and 
	signal to noise ratio, percentage of blue-only profiles, and red-only 
	profiles, the rate between them, and a description the observations: 
	temporal scan (TS, with the strip width expressed in granules), 
	large scan (LS), sunspot (SS), pore (P) and
         off the limb (OL).}} \label{latabla}\\

   \multicolumn{1}{c}{\textbf{Date Time}} &
   \multicolumn{1}{c}{\textbf{X scale }} &
   \multicolumn{1}{c}{\textbf{Y scale }} &
   \multicolumn{1}{c}{\textbf{( X, Y) }} &
   \multicolumn{1}{c}{\textbf{$\mu$}} &
   \multicolumn{1}{c}{\textbf{Exp. Time}} &
   \multicolumn{1}{c}{\textbf{S/N}} &
   \multicolumn{1}{c}{\textbf{B}} &
   \multicolumn{1}{c}{\textbf{R}} &
   \multicolumn{1}{c}{\textbf{B/R}} &
   \multicolumn{1}{c}{\textbf{Desc.}} \vspace{+0.05cm}\\
   \multicolumn{1}{c}{\textbf{}} &
   \multicolumn{1}{c}{\textbf{(\arcsec)}} &
   \multicolumn{1}{c}{\textbf{(\arcsec)}} &
   \multicolumn{1}{c}{\textbf{(\arcsec, \arcsec) }} &
   \multicolumn{1}{c}{\textbf{}} &
   \multicolumn{1}{c}{\textbf{(s)}} &
   \multicolumn{1}{c}{\textbf{}} &
   \multicolumn{1}{c}{\textbf{(\%)}} &
   \multicolumn{1}{c}{\textbf{(\%)}} &
   \multicolumn{1}{c}{\textbf{}} & 
   \multicolumn{1}{c}{\textbf{}} \\ [0.5ex] \hline
   \\[-1.8ex]
\endfirsthead

\multicolumn{3}{c}{{\tablename} \thetable{} -- Continued} \\[0.5ex]
   \multicolumn{1}{c}{\textbf{Date Time}} &
   \multicolumn{1}{c}{\textbf{X scale }} &
   \multicolumn{1}{c}{\textbf{Y scale }} &
   \multicolumn{1}{c}{\textbf{( X, Y) }} &
   \multicolumn{1}{c}{\textbf{$\mu$}} &
   \multicolumn{1}{c}{\textbf{Exp. Time}} &
   \multicolumn{1}{c}{\textbf{S/N}} &
   \multicolumn{1}{c}{\textbf{B}} &
   \multicolumn{1}{c}{\textbf{R}} &
   \multicolumn{1}{c}{\textbf{B/R}} &
   \multicolumn{1}{c}{\textbf{Desc.}}  \vspace{+0.05cm}\\ 
   \multicolumn{1}{c}{\textbf{}} &
   \multicolumn{1}{c}{\textbf{(\arcsec)}} &
   \multicolumn{1}{c}{\textbf{(\arcsec)}} &
   \multicolumn{1}{c}{\textbf{(\arcsec, \arcsec) }} &
   \multicolumn{1}{c}{\textbf{}} &
   \multicolumn{1}{c}{\textbf{(s)}} &
   \multicolumn{1}{c}{\textbf{}} &
   \multicolumn{1}{c}{\textbf{(\%)}} &
   \multicolumn{1}{c}{\textbf{(\%)}} &
   \multicolumn{1}{c}{\textbf{}} &
   \multicolumn{1}{c}{\textbf{}} \\ [0.5ex] \hline
   \\[-1.8ex]
\endhead

  \multicolumn{3}{l}{{Continued on Next Page\ldots}} \\
\endfoot

  \\[-1.8ex] 
\endlastfoot

2007.02.19 21:31 & 0.30 & 0.32 & (  -4.3,   7.6) & 1.000 &  1.6 &  1189 & 3.7 & 1.4 & 2.7 & TS (1) \\
2007.02.24 02:28 & 0.30 & 0.32 & ( 415.5, -70.3) & 0.899 &  1.6 &  1148 & 1.6 & 0.8 & 2.0 & TS (5) \\
2007.02.27 11:51 & 0.15 & 0.16 & (-271.2,  18.7) & 0.959 &  4.8 &   631 & 1.2 & 0.4 & 3.4 & TS (5) \& SS \\
2007.09.01 20:35 & 0.15 & 0.16 & (-153.1, 922.9) & 0.224 &  8.0 &   549 & 1.4 & 2.3 & 0.6 & LS  \\
2007.09.06 00:28 & 0.15 & 0.16 & (-153.1, 922.8) & 0.225 &  9.6 &   630 & 1.4 & 2.4 & 0.6 & LS  \\
2007.09.06 13:11 & 0.15 & 0.16 & (-214.5,   7.0) & 0.975 &  8.0 &  1146 & 3.5 & 1.8 & 1.9 & LS \\
2007.09.06 15:55 & 0.15 & 0.16 & ( -34.6,   7.0) & 0.999 &  8.0 &  1150 & 3.6 & 1.8 & 1.9 & LS \\
2007.09.07 11:44 & 0.15 & 0.16 & ( 145.4,   6.8) & 0.988 &  8.0 &  1138 & 3.4 & 1.8 & 1.8 & LS \\
2007.09.07 15:04 & 0.15 & 0.16 & ( 325.3,   6.8) & 0.941 &  8.0 &  1110 & 2.9 & 1.8 & 1.6 & LS \\
2007.09.09 07:05 & 0.15 & 0.16 & ( 646.8,   7.2) & 0.739 &  9.6 &   957 & 2.1 & 2.8 & 0.7 & LS \\
2007.09.09 13:05 & 0.15 & 0.16 & ( -79.0,-892.6) & 0.359 &  9.6 &   868 & 1.9 & 2.9 & 0.7 & LS \\
2007.09.10 01:15 & 0.15 & 0.16 & (-153.1, 922.9) & 0.225 &  9.6 &   634 & 1.4 & 2.5 & 0.5 & LS \\
2007.09.10 08:00 & 0.15 & 0.16 & ( -34.2,   6.8) & 0.999 &  9.6 &  1275 & 3.6 & 1.8 & 2.0 & LS \\
2007.09.11 06:10 & 0.15 & 0.16 & (-935.5,-291.0) & 0.000 & 12.8 &   950 & 1.7 & 2.9 & 0.6 & LS \\
2007.09.11 14:06 & 0.15 & 0.16 & (-153.1, 912.7) & 0.266 &  9.6 &   682 & 1.4 & 2.5 & 0.5 & TS (1) \\
2007.09.15 12:44 & 0.15 & 0.16 & ( -92.0,-173.1) & 0.979 &  9.6 & 14438 & 3.4 & 1.8 & 1.8 & LS \\
2007.09.16 07:23 & 0.15 & 0.16 & ( -21.8,   6.5) & 1.000 &  4.8 &   873 & 3.6 & 1.5 & 2.4 & TS (1) \\
2007.09.19 07:11 & 0.15 & 0.16 & ( -16.6, 456.7) & 0.879 &  3.2 &   685 & 3.4 & 1.3 & 2.7 & TS (2) \\
2007.09.24 11:56 & 0.15 & 0.16 & (  -6.9,   7.4) & 1.000 &  1.6 &   544 & 3.6 & 0.9 & 4.0 & TS (1) \\
2007.09.24 19:32 & 0.15 & 0.16 & ( -15.8,   6.7) & 1.000 & 12.8 &  1461 & 3.1 & 1.9 & 1.7 & LS \\
2007.09.24 20:42 & 0.15 & 0.16 & (  -5.1,   6.7) & 1.000 & 12.8 &  1489 & 3.4 & 1.8 & 1.9 & LS \\
1007.09.24 21:52 & 0.15 & 0.16 & (   5.7,   7.0) & 1.000 & 12.8 &  1464 & 3.2 & 1.9 & 1.7 & LS \\
2007.09.25 12:59 & 0.15 & 0.16 & ( -26.3,   6.5) & 1.000 &  1.6 &   547 & 4.5 & 1.3 & 3.5 & TS (1) \\
2007.09.26 08:15 & 0.15 & 0.16 & (  -7.7,   6.4) & 1.000 &  1.6 &   531 & 3.6 & 1.1 & 3.4 & TS (1) \\
2007.09.26 10:45 & 0.15 & 0.16 & (  -4.9,   6.5) & 1.000 &  1.6 &   547 & 2.9 & 0.9 & 3.3 & TS (1) \\
2007.09.26 18:03 & 0.15 & 0.16 & ( 835.6,   7.6) & 0.492 & 12.8 &   768 & 1.5 & 2.5 & 0.6 & LS \\
2007.09.27 01:01 & 0.15 & 0.16 & (-1004.,   7.5) & 0.000 & 12.8 &   887 & 1.5 & 2.6 & 0.6 & LS \& OF \\
2007.09.28 07:00 & 0.15 & 0.16 & ( -19.3,   6.4) & 1.000 &  1.6 &   535 & 3.2 & 1.1 & 2.8 & TS (1) \\
2007.09.29 06:51 & 0.15 & 0.16 & ( -20.7,   6.4) & 1.000 &  1.6 &   539 & 4.2 & 1.3 & 3.1 & TS (1) \\
2007.10.01 08:21 & 0.15 & 0.16 & (  -6.6,   7.3) & 1.000 &  1.6 &   535 & 4.2 & 1.3 & 3.2 & TS (1) \\
2007.10.04 10:41 & 0.30 & 0.32 & ( -16.1,   6.5) & 1.000 &  1.6 &  1169 & 4.0 & 1.4 & 2.9 & TS (10) \\
2007.10.05 08:05 & 0.30 & 0.32 & ( -17.6,   6.5) & 1.000 &  1.6 &  1139 & 4.1 & 1.2 & 3.4 & TS (10) \\
2007.10.05 11:23 & 0.30 & 0.32 & ( -15.4,   6.3) & 1.000 &  1.6 &  1167 & 4.3 & 1.9 & 2.3 & TS (10) \\
2007.10.18 19:11 & 0.30 & 0.32 & ( 434.5,-221.7) & 0.861 &  1.6 &  1085 & 3.3 & 1.7 & 1.9 & TS (8) \\
2007.10.23 05:28 & 0.30 & 0.32 & (-121.9,-250.4) & 0.957 &  1.6 &  1141 & 3.7 & 1.5 & 2.4 & TS (8) \\
2007.10.24 01:04 & 0.30 & 0.32 & (  58.3,-249.5) & 0.964 &  1.6 &  1147 & 3.3 & 1.2 & 2.8 & TS (8) \& P\\
2007.10.25 00:02 & 0.30 & 0.32 & ( 546.6,-342.5) & 0.741 &  1.6 &  1066 & 3.6 & 1.8 & 1.4 & TS (8) \\
2007.10.26 03:51 & 0.30 & 0.32 & ( -34.6,   6.3) & 0.999 &  1.6 &  1158 & 4.5 & 1.4 & 3.1 & TS (8) \\
2007.10.26 19:50 & 0.30 & 0.32 & ( -53.3,   6.0) & 0.998 &  1.6 &  1138 & 4.1 & 1.4 & 3.0 & TS (8) \\
2007.10.27 02:12 & 0.15 & 0.16 & ( 743.1,   5.1) & 0.633 &  1.6 &   484 & 2.7 & 1.8 & 1.5 & TS (2) \\
2007.10.29 16:09 & 0.15 & 0.16 & ( -25.5,   6.3) & 1.000 &  1.6 &   539 & 4.3 & 1.2 & 3.7 & TS (2) \\
2007.10.29 17:54 & 0.15 & 0.16 & (  -9.1,   6.4) & 1.000 &  1.6 &   543 & 4.8 & 1.2 & 3.9 & TS (2) \\
2007.10.29 19:31 & 0.15 & 0.16 & (   6.1,   6.4) & 1.000 &  1.6 &   545 & 3.5 & 1.1 & 3.2 & TS (2) \\
2007.10.29 21:08 & 0.15 & 0.16 & (  21.3,   6.6) & 1.000 &  1.6 &   545 & 4.0 & 0.9 & 4.6 & TS (2) \\
2008.07.22 14:05 & 0.30 & 0.32 & ( -22.1,   7.5) & 1.000 &  1.6 &  1135 & 4.2 & 1.5 & 2.7 & TS (2) \\
2008.07.24 15:01 & 0.30 & 0.32 & ( -13.6,   7.8) & 1.000 &  1.6 &  1136 & 4.4 & 1.8 & 2.5 & TS (2) \\
2008.07.28 14:13 & 0.30 & 0.32 & ( -20.9,   7.6) & 1.000 &  1.6 &  1136 & 3.6 & 1.5 & 2.4 & TS (2) \\
2008.07.28 15:52 & 0.30 & 0.32 & (  -6.3,   7.6) & 1.000 &  1.6 &  1134 & 3.5 & 1.2 & 2.8 & TS (2) \\
2008.07.30 14:22 & 0.30 & 0.32 & (  65.9, -54.0) & 0.996 &  1.6 &  1135 & 3.1 & 0.9 & 3.4 & TS (2) \\
2008.07.31 14:25 & 0.30 & 0.32 & (  -8.1, 867.5) & 0.428 &  1.6 &   886 & 1.9 & 3.0 & 0.6 & TS (2) \\
2008.07.31 16:20 & 0.30 & 0.32 & (  -8.1, 867.4) & 0.428 &  1.6 &   884 & 1.9 & 3.3 & 0.6 & TS (2) \\
2008.12.22 00:00 & 0.15 & 0.16 & ( -79.3,  -2.1) & 0.997 &  4.8 &   843 & 3.1 & 1.1 & 2.8 & TS (5) \\
2008.12.23 00:00 & 0.15 & 0.16 & ( 148.1,  -0.1) & 0.988 &  4.8 &   837 & 3.5 & 1.6 & 2.2 & TS (5) \\
2008.12.25 11:00 & 0.30 & 0.32 & (-241.2,  67.5) & 0.965 &  1.6 &   744 & 4.1 & 1.5 & 2.8 & LS  \\
2008.12.25 12:00 & 0.30 & 0.32 & (-231.9,  67.6) & 0.968 &  1.6 &   752 & 4.2 & 1.2 & 3.4 & LS  \\
2008.12.25 13:00 & 0.30 & 0.32 & (-222.6,  67.8) & 0.970 &  1.6 &   753 & 4.1 & 1.2 & 3.3 & LS  \\
2008.12.25 14:00 & 0.30 & 0.32 & (-213.2,  67.8) & 0.972 &  1.6 &   755 & 3.9 & 1.3 & 3.0 & LS  \\
2008.12.25 15:00 & 0.30 & 0.32 & (-203.9,  67.7) & 0.975 &  1.6 &   751 & 3.7 & 1.4 & 2.7 & LS  \\
2008.12.25 16:00 & 0.30 & 0.32 & (-194.5,  68.2) & 0.977 &  1.6 &   751 & 3.9 & 1.2 & 3.2 & LS  \\
2008.12.25 17:00 & 0.30 & 0.32 & (-185.1,  68.3) & 0.979 &  1.6 &   753 & 3.9 & 1.5 & 2.6 & LS  \\
2008.12.25 21:01 & 0.30 & 0.32 & (-147.1,  68.9) & 0.986 &  1.6 &   749 & 4.2 & 1.6 & 2.7 & LS  \\
2008.12.25 22:01 & 0.30 & 0.32 & (-166.2,  69.0) & 0.982 &  1.6 &   713 & 4.2 & 1.4 & 3.0 & LS  \\
2009.05.28 10:13 & 0.30 & 0.32 & ( 234.0,-516.5) & 0.807 &  1.6 &  1044 & 2.0 & 1.1 & 1.9 & TS (5)  \\
2009.05.29 09:11 & 0.30 & 0.32 & ( 382.8,-519.0) & 0.741 &  1.6 &  1010 & 2.3 & 1.9 & 1.2 & TS (5) \\
2009.05.31 08:47 & 0.30 & 0.32 & (-650.7, 393.6) & 0.610 &  1.6 &   929 & 1.8 & 2.1 & 0.9 & TS (5) \\
2009.06.01 09:25 & 0.30 & 0.32 & (-520.9, 394.4) & 0.733 &  1.6 &   939 & 0.7 & 0.4 & 1.7 & TS (8) \& SS \\
2009.06.02 10:03 & 0.30 & 0.32 & (-331.0, 393.9) & 0.844 &  1.6 &   838 & 0.7 & 0.6 & 1.2 & TS (8) \& SS \\
2009.06.03 07:54 & 0.30 & 0.32 & (-165.1, 432.6) & 0.876 &  1.6 &   990 & 1.4 & 1.2 & 1.2 & TS (5)  \& P\\
2009.06.03 09:01 & 0.30 & 0.32 & (-155.9, 432.6) & 0.878 &  1.6 &   964 & 1.4 & 1.3 & 1.0 & TS (5) \\
2009.06.05 08:39 & 0.30 & 0.32 & ( 220.0, 428.7) & 0.865 &  1.6 &  1055 & 1.9 & 1.4 & 1.4 & TS (5) \\
2009.06.06 09:16 & 0.30 & 0.32 & ( 402.9, 370.6) & 0.821 &  1.6 &  1040 & 1.3 & 0.7 & 1.7 & TS (8) \\
2009.06.07 08:16 & 0.30 & 0.32 & ( 560.4, 368.9) & 0.715 &  1.6 &   950 & 1.5 & 1.5 & 1.0 & TS (8) \\
2009.06.07 09:54 & 0.30 & 0.32 & ( 570.7, 368.8) & 0.706 &  1.6 &   993 & 1.4 & 1.6 & 0.9 & TS (8) \\
\hline
\end{longtable}
\end{center}

\end{document}